\documentclass[preprint,eqsecnum,showpacs,preprintnumbers,amsmath,amssymb]{revtex4}

\usepackage{graphicx}
\usepackage{dcolumn}
\usepackage{bm}

\begin{document}

\preprint{Taraskin, Spectral properties ... }  
  
\title{Spectral properties of disordered fully-connected graphs}

\author{S.~N.~Taraskin}  
 \email{snt1000@cam.ac.uk}  
\affiliation{St. Catharine's College and Department of Chemistry, University of Cambridge,  
             Cambridge, UK}

\date{\today}
  
\begin{abstract}  
The spectral properties of disordered fully-connected graphs 
with a special type of the node-node interactions are investigated. 
The approximate analytical expression for the ensemble-averaged 
spectral density for the Hamiltonian defined on 
the fully-connected graph is derived and analysed both for the electronic 
and vibrational problems which can be related to the contact process and 
to the problem of stochastic diffusion, respectively. 
It is demonstrated how to evaluate the extreme eigenvalues and use them 
for finding the  lower bound estimates of the critical parameter for the 
contact process on the disordered fully-connected graphs.   
\end{abstract}  
  
\pacs{
 63.50.+x, 71.23.-k, 02.10.Ox, 89.75.Hc
}
 
  
\maketitle  
  
 
\section{Introduction} 
\label{s0}

The spectral properties of complex networks are
 of great current interest, both for practical applications
and from a fundamental point of view 
\cite{Dorogovtsev_02,Albert_02,Dorogovtsev_03:book,Fortin_05,Iguchi_05,Iguchi_05:exact,Dorogovtsev_03,Hastings_03,Farkas_01,Goh_01,Monasson_99,Biroli_99}. 
The physical phenomena occurring in the network can be described using the 
operators defined for the network. 
For example, the Hamiltonian describes electronic excitations in the 
network of atoms (nodes) characterized by energy levels 
communicating with each other by hopping integrals (links). 
The Laplacian operator for a set of atoms connected by elastic 
springs describes vibrational excitations \cite{Maradudin_71} 
or transport phenomena, e.g. stochastic diffusion 
of random-walk type \cite{Bottger_85:book,Rudnick_04:book}. 
The Liouville operator characterizes the spread of epidemics in the network 
\cite{Liggett_85:book,Marro_99:book} and the 
connectivity operator gives knowledge of the network topology 
\cite{Dorogovtsev_02,Albert_02,Dorogovtsev_03:book}. 
In the  matrix representation, these operators are characterized by matrices,  
the eigenspectrum of which gives  rather 
complete information of e.g. dynamical properties of the network. 
The main difficulty in analytical evaluation of the network spectrum is 
in an inherent disorder incorporated into the network matrices. 
This can be topological disorder related to irregular Euclidean 
arrangements of the nodes (e.g. atoms in liquids and glasses) 
and/or disorder in connectivity 
(complex networks e.g. of scale-free or small-world type 
\cite{Dorogovtsev_03:book}) or disorder in 
parameters associated with the nodes and links 
(e.g. mass and force-constant 
disorder for the vibrational problem on a regular lattice 
\cite{Taraskin_02:JPCM} or 
substitutional disorder in metallic alloys \cite{Ehrenreich_76}). 

The first task of the spectral analysis is to evaluate the spectral 
density of the relevant operator for a particular realization of 
disorder. 
For an ordered system, all the realizations are identical (zero disorder) 
and the spectrum of at least some operators can be easily found. 
For disordered systems, 
this is a highly non-trivial problem and can be solved implicitly 
only for some simple models of local perturbation in a reference 
not necessarily ordered system (see Ref.~\cite{Bogomolny_01} and 
references therein).  
The next task is to perform the averaging over different realizations 
of disorder (configurational averaging). 
The configurationally averaged spectral density can be used then for 
 comparison with the experimentally measured spectrum (e.g. by inelastic 
neutron scattering for vibrational excitations \cite{Maradudin_71}) 
or with other observable characteristics (thermodynamical values such 
as heat capacity or linear response functions, e.g. the dielectric response 
function). 
Such a comparison with experimental observables makes sense if the 
measurable quantity is self-averaging, i.e. if the difference between the mean 
value of the observable for a particular realization of disorder in a 
macroscopically large system and the cofigurationally averaged value of this 
observable tends to zero with increasing system size 
\cite{Binder_86,Kramer_93}.  
Some of the observables such as thermodynamical characteristics of spin 
glasses away from the phase transition points 
and of normal glasses at not very low temperatures are self-averaging. 
However, some of the observables such as thermodynamical characteristics 
of spin glasses at criticality  \cite{Wiseman_98}, low temperature 
electron conductance \cite{Kramer_93}and  dielectric response in disordered 
semiconductors with strong disorder \cite{Kawasaki_03} are sample-dependent 
and thus are not self-averaging. 

The spectral density of at least Hamiltonians and Laplacians 
exhibits decreasing  fluctuations  with increasing system size 
(even at the localization/delocalization threshold point) 
and thus is expected to be a self-averaging characteristic 
(see e.g. \cite{Rossum_94}). 
This has been seen implicitly in numerous computer simulations of spectral 
properties of disordered systems 
(see e.g. \cite{Schreiber_96:book} and references therein) and 
in experiment \cite{Elliott_90:book}.     
Usually, the ensemble averaging is tackled by means of 
mean-field theories \cite{Mirlin_00:review,Guhr_98:review} 
with possible use of the replica method \cite{Dotsenko_00:book} or introducing 
supersymmetry \cite{Efetov_83:review,Rossum_94} 
 but in some cases the ''exact'' solutions are available. 
They are quite rare and the examples are the semicircular spectrum 
for the fully-connected graph (FCG) with random normally distributed 
node-node interactions 
\cite{Brody_81,Mehta_91:book} (see also Ref.~\cite{Pastur_72}) 
and  Lloyd's model  for a special type of the on-site  
disorder  and any network topology \cite{Lloyd_69,Lehmann_73}. 

The main aim of this paper is to present a model 
dealing with the disordered Hamiltonians defined on the FCG 
with a special type 
of the node-node interactions and a rather general type of the 
on-site characteristics. 
It is possible to find an implicit analytical expression for the spectral 
density for a particular realization of disorder and then 
to perform analytically 
the ensemble averaging of the spectrum of the Hamiltonian   
with precision up to $O(N^{-1})$ with $N$ being the number 
of nodes in the FCG and thus to demonstrate the self-averaging 
properties of the spectral density. 
The analytical results for the ensemble-averaged 
spectral density  are available due to 
the existence of an exact solution  for the matrix elements of 
the resolvent operator  for a particular realization of disorder. 
The general solution is specified and analysed for several 
particular problems including the electronic and vibrational problems 
with multiplicative interactions defined on the FCG. 
The electronic problem is equivalent to the contact process in the dilute 
regime and the results can be used for the lower bound estimate of 
the critical point for the contact process which describes e.g. the spread 
of epidemics through the network. 
The vibrational problem is equivalent to the stochastic transport 
problem and the results can be used for the investigation of dynamics 
of information packets propagating through a communication network.  
The main analytical results both for the electronic and 
vibrational problems  are supported by direct 
numerical diagonalization of the Hamiltonian. 
  
The paper is organized in the following manner. 
The formulation of the problem is given in Sec.~\ref{s1}. 
Several simple examples are considered in Sec.~\ref{s2} followed 
by the general  solution of the problem for polynomial interactions
in Sec.~\ref{s3}. 
The limitations of the approach are discussed 
in Sec.~\ref{s4}. 
The conclusions are made in Sec.~\ref{s5} and  
some derivations are presented in Appendices~\ref{app_a} and~\ref{app_b}. 

\section{Formulation of the problem} 
\label{s1}

Let us consider a FCG containing $N$ nodes. 
Each node $i$ is characterized 
by the parameter $\varepsilon_i$ (node bare energy)  
and the link between nodes $i$ and $j$ 
by parameter $V_{ij}$ (node-node interaction). 
Then we define an operator $\hat{\bf H}$ (''Hamiltonian'') on this FCG in the 
following manner, 
\begin{equation}
\hat{\bf H} = \sum_i \left(\epsilon_i + 
\gamma \sum_{j\ne i} V_{ij} + V_{ii}\right)|i\rangle\langle i| 
- \sum_{i,j}V_{ij}|i\rangle\langle j|
~,
\label{e1_1}
\end{equation}
where the self-interaction  matrix element $V_{ii}$ 
is introduced for convenience (the Hamiltonian, 
in fact, does not depend on it). 
The tuning parameter $\gamma$ gives an opportunity to distinguish 
between two types of problems: (i) electronic-like for $\gamma=0$ and 
(ii) vibrational-like when $\gamma=1$ and all $\epsilon_i=0$ 
(see also \cite{Mezard_99}).   
For vibrational problem, the operator $\hat{\bf H}$ is the Hessian operator 
and its elements obey the sum rule, $V_{ii}=\sum_{j\ne i}V_{ij}$, which 
follows from 
the global translational invariance of the Hamiltonian \cite{Maradudin_71}. 

Both the electronic and vibrational problems  are usually defined 
on networks with Euclidean topology describing real materials. 
Below, we consider a FCG and thus the physical meaning of the 
Hamiltonian~(\ref{e1_1}) defined on the FCG should be specified. 
This can be done by introducing two mappings. 

First, the electronic Hamiltonian ($\gamma=0$) is equivalent to the 
Liouville operator, $\hat{\mathcal L}$, describing the contact process 
in the dilute regime \cite{Taraskin_05:PRE_CP}. 
Indeed, the time evolution of the state vector, $|P(t)\rangle$, 
for the contact process  is governed by the 
master equation describing the conserved probability 
flow \cite{Marro_99:book}, 
$\partial_t |P(t)\rangle = \hat{\mathcal L} |P(t)\rangle
$, which can be rewritten in the dilute regime as 
(see Ref.~\cite{Taraskin_05:PRE_CP} for more detail)  
\begin{equation}
\partial_t \overline{P}_{i}(t) = 
-r_i \overline{P}_{i}(t) + \sum_{j\ne i} 
W_{ji}\overline{P}_{j}(t) 
~,  
\label{e1_1b}
\end{equation}
where $\overline{P}_{i}(t)$ is the probability of finding node $i$ 
in an occupied  state independent of the occupation of all the other nodes 
which can be in two states, occupied (infected) or unoccupied 
(susceptible), $r_i$ is the recovery rate for node $i$ and 
$W_{ji}$ is the transmission (infection) rate between node $j$ and $i$. 
The formal solution of Eq.~(\ref{e1_1b}) is given by 
\begin{equation}
|\overline{P}(t)\rangle = e^{\hat{\mathcal L}t} |\overline{P}(0)\rangle
= \sum_{j}e^{\varepsilon_j t}\langle {\bf e}^j|\overline{P}(0)\rangle 
|{\bf e}^j \rangle~,
\label{e1_1c}
\end{equation}
with $\varepsilon_j$ and $|{\bf e}^j \rangle$ being the eigenvalues and 
eigenvectors of the Liouville operator, respectively, which 
coincides with the Hamiltonian~(\ref{e1_1}) for $\gamma=0$, $r_i=-\epsilon_i$ 
and $V_{ij}=-W_{ij}$. 
The long-time behaviour of the contact process in the dilute regime 
 is defined by the maximum eigenvalue, $\varepsilon_{j,{\text{max}}}$, and if 
$\varepsilon_{j,{\text{max}}} < 0$ then the epidemic goes to extinction and 
it invades if $\varepsilon_{j,{\text{max}}} > 0$. 
The approximate rate equation~(\ref{e1_1b}) has been obtained 
by replacing the term 
$(\overline{P}_j(t)- \overline{P}_{ji}(t))$ 
(where $\overline{P}_{ji}(t)$ is the probability for both nodes $i$ and $j$ 
to be occupied independent of the state of all the other nodes) 
in the exact equation with $\overline{P}_j(t)$. 
Such an approximation enhances the transmission of the disease and thus 
the estimate of the critical point obtained from the solution of the 
equation, $\varepsilon_{j,{\text{max}}} = 0$, gives a reliable lower bound 
estimate of the critical parameter for the contact process meaning 
that if the disease does not spread in the dilute regime then it certainly 
does not spread in the system. 
This can be practically important   
for controlling  epidemics in disordered systems where the estimate of the 
exact value of the critical parameter is a rather complicated task 
\cite{Hooyberghs_03,Hooyberghs_04,Vojta_04}. 
    
The second mapping connects the vibrational problem to the problem of 
stochastic diffusion through a net. 
While the contact process describes a propagation of excitations 
(infected nodes) through the net with a not-conserved number of excitations 
(the number of infected nodes changes with time) then the standard stochastic 
diffusion deals with propagation of the conserved number of excitations 
(diffusing particles) through the net by means of diffusional jumps 
(characterized by the rates $W_{ij}$) between the nodes. 
The balance equation for stochastic diffusion coincides with 
Eq.~(\ref{e1_1b}) where $r_i$ is replaced by $\sum_{j\ne i} W_{ij}$ 
\cite{Bottger_85:book,Webman_81,Odagaki_81} which reflects the conservation 
of the number of particles. 
Therefore, under the assumption of symmetric transition rates, 
$W_{ij}=W_{ji}$, stochastic diffusion through the network is 
described by the Hamiltonian~(\ref{e1_1}) with $\gamma=1$, $\epsilon_i=0$ and 
$V_{ij}=-W_{ij}$. 
For complex networks, such as the FCG, the diffusing particles can 
be associated e.g. with the information packets propagating through the 
communication network \cite{Dorogovtsev_03:book}. 
The quantity of interest can be e.g. the return probability of the 
diffusing particle to the starting place, 
$\langle P_0(t) \rangle = N^{-1}\text{Tr}\,\exp\{\hat{\bf H}t\}$ 
(see e.g. \cite{Friedrichs_88}). 

The aim of our analysis is to find the eigenspectrum 
of Hamiltonian~(\ref{e1_1}) defined on the FCG. 
In the site (node) basis, the Hamiltonian matrix is fully 
dense and, for the general case of arbitrary parameters $\epsilon_i$ and 
$V_{ij}$, its diagonalization is not a trivial task and the solution 
is not currently known. 
However, for the classical case of a random matrix belonging to the 
Gaussian Orthogonal Ensemble, when the off-diagonal (diagonal) 
elements are independent 
and normally distributed (with variance doubled for diagonal elements), 
the configurationally averaged spectrum of semicircular shape can 
be evaluated analytically (with errors of $O(N^{-1})$) 
\cite{Mehta_91:book,Guhr_98:review}.  
One of the key features of the matrices belonging to the 
Gaussian orthogonal ensemble is the statistical 
independence of the matrix elements. 

Below, we suggest another class of real symmetric matrices for 
which the spectral density 
can be found for a particular realization of disorder and then 
the ensemble averaging can be performed analytically. 
These are matrices with a particular (polynomial) type of the node-node 
interactions: 
\begin{equation}
V_{ij}=\bm{\varphi}_i^T \bm{\alpha} \bm{\varphi}_j
~,
\label{e1_2}
\end{equation}
where $\bm{\varphi}_i^T$ is a $n$-dimensional row-vector, 
$\bm{\varphi}_i^T = (1,\phi_i,\phi_i^2,\ldots,\phi_i^{n-1})$ and 
$\bm{\alpha}$ is a real symmetric $n\times n$ matrix of interaction 
coefficients, so that $V_{ij}(\phi_i,\phi_j)$ is a symmetric 
polynomial form of order $2n-2$ with respect to $\phi_i$ and $\phi_j$, 
e.g. $V_{ij}=\alpha_{11} + \alpha_{12}(\phi_i +\phi_j) + 
\alpha_{22}\phi_i\phi_j$ for $n=2$. 
The values of $\phi_i$ are independent random variables characterized, in 
general, by different probability distribution functions, 
$\rho_{\phi_i}(\phi_i)$. 
The order of interaction, $n$, is supposed to be much less 
than the number of nodes in the system, $n\ll N$. 
The diagonal elements, $\varepsilon_i$, in such matrices 
are also random independent variables 
distributed according to the probability distribution functions, 
$\rho_{\epsilon_i}(\epsilon_i)$. 
The probability distributions $\rho_{\phi_i}(\phi_i)$ and  
$\rho_{\epsilon_i}(\epsilon_i)$ are assumed to have all finite moments 
unless it is stated differently. 

We demonstrate below that the matrix elements of the resolvent 
(Green's function) operator, 
$\hat{\bf G}$, defined by the equation, 
$(\varepsilon {\hat{\bf I}} - \hat{\bf H})\hat{\bf G} = {\hat{\bf I}}$, 
can be found exactly for the interactions given by Eq.~(\ref{e1_2}). 
This means, that the density of states (DOS), $g(\varepsilon)$, 
is available for a particular realization of disorder according to 
the following identity  \cite{Economou_83:book},  
\begin{equation}
g(\varepsilon)=-\frac{1}{\pi N} \text{Im}~\text{Tr}\,
{\hat{\bf G}}(\varepsilon+\text{i}0) 
~.
\label{e1_4}
\end{equation}
Due to the availability of the analytical expression for $g(\varepsilon)$, its 
configurational averaging, $\langle g(\varepsilon)\rangle $,   
can also be undertaken analytically by means of the following integration 
(for $N\to\infty$),
\begin{equation}
\langle g(\varepsilon)\rangle =
\left\langle N^{-1}\sum_i \delta(\varepsilon-\varepsilon_i)\right\rangle 
\equiv 
\idotsint N^{-1}\sum_i \delta(\varepsilon-\varepsilon_i) 
\prod_i \rho_{\phi_i}(\phi_i) \rho_{\epsilon_i}(\epsilon_i)\, 
\text{d}\phi_i\text{d}\epsilon_i
~,
\label{e1_3}
\end{equation}
where $\varepsilon_i$ are the eigenvalues (eigenenergies) of the Hamiltonian.

\section{Simple examples} 
\label{s2}

Before considering the general case of the polynomial node-node 
interactions~(\ref{e1_2}), we present, first, three simple examples 
where the exact solution of the problem is available. 
These examples are for (i) the ideal FCG, (ii) 
the binary FCG and (iii) Lloyd's model defined 
on the FCG.
In all cases, the biological (epidemiological) applications 
of the results are discussed.  

\subsection{Ideal fully-connected graph} 
\label{s2_1}

\begin{figure}[ht] 
\vskip2truecm
\centerline{\includegraphics[width=13cm]{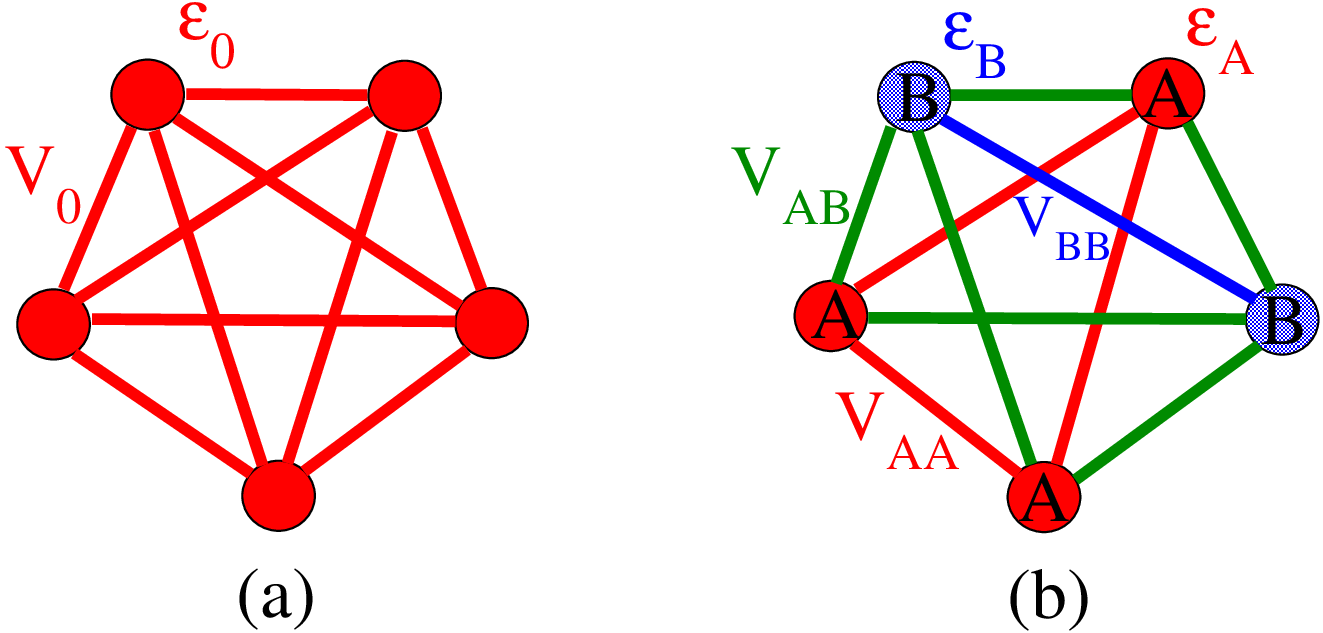}}  
\caption{(Color online)
Ideal (a) and binary (b) fully-connected graphs with $N=5$. 
  }  
\label{f1} 
\end{figure}  

A trivial case we need as a reference for further analysis is an ideal FCG, 
for which all the on-site energies are identical, 
$\rho_{\epsilon_i}(\epsilon_i)=\delta(\epsilon_i - \epsilon_0)$, 
and all the interactions are the same, 
$\rho_{\phi_i}(\phi_i)=\delta(\phi_i - \phi_0)$ 
(so that $V_{ij}= V_0$), see Fig.~\ref{f1}(a). 
The diagonal elements of the resolvent are, 
\begin{equation}
G_{ii}^{(0)}=\frac{N-1}{\varepsilon -\tilde\epsilon_0} 
+\frac{1}{\varepsilon -\tilde{\epsilon}_0+NV_0}  
~,
\label{e2_1_1}
\end{equation}
where $\tilde{\epsilon}_0=\epsilon_0+V_0 +\gamma(N-1)V_0$, and thus 
the spectrum of the ideal FCG contains two delta-functions, one of them is 
$(N-1)$-degenerate, 
\begin{equation}
g_0(\varepsilon)=\langle g_0(\varepsilon)\rangle =
\left(1-\frac{1}{N}\right)\delta
\left(\varepsilon-\tilde{\epsilon}_0\right) 
+ 
\frac{1}{N}\delta\left(\varepsilon-\tilde{\epsilon}_0+NV_0\right) 
~. 
\label{e2_1_2}
\end{equation}
The spectrum of the Hamiltonian defined on the ideal FCG 
obviously obeys the ''energy-conservation'' principle, 
$\text{Tr}\,{\hat{\bf H}}= N\epsilon_0$, meaning that the interactions 
do not change the total bare energy. 
If disorder is introduced in the ideal system it is quite natural to 
expect the broadening of the $(N-1)$-degenerate level to the band and possibly 
the appearance of more levels split from the band. 
This is exactly what happens in the disordered FCG according to 
the analysis presented below.   

In terms of biological applications, Eq.~(\ref{e2_1_2}) can be used 
for estimating the value of the critical parameter $\eta_c$ 
(this can be  the ratio of the typical transmission and recovery rates) 
 separating the absorbing ($\eta < \eta_c$) and active ($\eta > \eta_c$) 
states of the system with respect to the spread of the contact process 
(epidemic). 
Indeed, the maximum eigenvalue of the spectrum  coincides with the 
position of the non-degenerate $\delta$-function and is equal to  
$\varepsilon_{\text{max}}=(N-1)W_0 - r_0$ (bearing in mind 
that $\gamma=0$, $W_0=-V_0$ and $r_0=-\epsilon_0$). 
The solution of the equation, $\varepsilon_{\text{max}}=0$, gives 
a standard mean-field estimate for the critical parameter 
$\eta_c^* = (W_0/r_0)_c^* = (N-1)^{-1}$ \cite{Marro_99:book}. 
Obviously, if the transmission rate $W_0$ is $N$-independent then the 
critical value $\eta_c^*$ approaches zero for large values of $N$ and 
the system is always in the active state \cite{Dorogovtsev_03:book}. 
However, if we assume that the transmission rate is inversely 
proportional to the number of nodes (in a migrating biological population 
the interaction time between the members of the population can be inversely 
proportional to the population size), $W_0=w_0/N$, with $w_0$ being 
independent of $N$, then the critical point exists and the estimate 
for the critical parameter is $(w_0/r_0)^*_c = 1$.

\subsection{Binary fully connected graph} 
\label{s3_5}

Another simple example of the network is a binary FCG (see Fig.~\ref{f1}(b)), 
 for which the spectrum for a particular realization of disorder 
is available and  configurational 
averaging of the DOS can be performed exactly. 
The binary FCG consists of nodes of two types, $A$ and $B$. 
The on-site energies $\epsilon_A$ and $\epsilon_B$ and the 
node-node interactions $V_{AA}$, $V_{BB}$ and $V_{AB}$ are defined 
by the types of the nodes. 
The node-node interactions, in general, are not of multiplicative form  and 
can be described by Eq.~(\ref{e1_2}) only if $V_{AB}^2=V_{AA}V_{BB}$. 
The only random parameter for the binary FCG is the number of nodes of 
a certain type, e.g. $N_{A}$, which is defined by the probability 
$p$ (parameter of the model) for a node to be of type $A$. 
The values of $N_A$ or equivalently of concentration $c=N_A/N$ are 
distributed according to the binomial probability distribution, 
$\rho_c(c)$ with the expectation value $E[c]=p$ and variance 
$\text{Var}[c]=p(1-p)N^{-1}$ which is close to the variance of the normal 
distribution for $N\to\infty$. 

The spectral density for the electronic ($\gamma=0$) 
Hamiltonian~(\ref{e1_1}) defined on the binary FCG 
characterized by a  particular value of concentration  $c$ 
contains four $\delta$-functions, 
\begin{equation}
g(\varepsilon) = (c-N^{-1})\delta(\varepsilon-\tilde{\epsilon}_A)+ 
(1-c-N^{-1})\delta(\varepsilon-\tilde{\epsilon}_B) 
+N^{-1}\sum_{i=1,2}\delta(\varepsilon - \varepsilon_i)
~, 
\label{e3_13}
\end{equation}
where $\tilde{\epsilon}_A=\epsilon_A+V_{AA}$ and 
$\tilde{\epsilon}_B=\epsilon_B+V_{BB}$. 
Only the two last $\delta$-functions, 
\begin{equation}
g_1(\varepsilon) = N^{-1}\sum_{i=1,2}\delta(\varepsilon - \varepsilon_i)
~, 
\label{e3_13a}
\end{equation}
depend on the random parameter $c\in [0,1]$ and thus should 
be configurationally averaged. 
The values of $\varepsilon_i$ in Eq.~(\ref{e3_13a}) are the roots of the 
spectral determinant, $D(\varepsilon_i)=0$, where 
\begin{equation}
D(\varepsilon)=
\left(1+\frac{cNV_{AA}}{\varepsilon-\tilde{\epsilon}_A} \right)
\left(1+\frac{(1-c)NV_{BB}}{\varepsilon-\tilde{\epsilon}_B} \right) - 
\frac{c(1-c)N^2V_{AB}^2}{(\varepsilon-\tilde{\epsilon}_A)
(\varepsilon-\tilde{\epsilon}_B)}
~. 
\label{e3_14}
\end{equation}
The above expression for $D(\varepsilon)$ can be derived in a manner 
similar to the derivation 
for multiplicative interactions (see Appendix~\ref{app_a}). 

For simplicity, we consider a symmetric binary FCG characterized 
by $\epsilon_A=\epsilon_B=\epsilon_0=-r_0$ and 
$V_{AA}=V_{BB}=V_0$ and also assume that the node-node interactions 
are negative and inversely proportional to $N$, so that the interaction 
parameters, $w=-NV_0\,$ and $\lambda = -NV_{AB}$, are positive and 
$N$-independent. 
In this case, the $c-\varepsilon$ map defined by the 
equation, $D(\varepsilon)=0$, with $D(\varepsilon)$ obeying 
Eq.~(\ref{e3_14}), 
is given by the following bilinear form, 
\begin{equation}
(\varepsilon+r_0 -w/2)^2 +(\lambda^2-w^2)(c-1/2)^2=\lambda^2/4
~,
\label{e3_15}
\end{equation}
where we have ignored the terms $\propto N^{-1}$. 
Therefore, the positions of the $\delta$- functions in 
Eq.~(\ref{e3_13a}) can be 
found as the roots of Eq.~(\ref{e3_15}) for a particular value of $c$, 
\begin{equation}
\varepsilon_{1,2}=-r_0+\frac{w}{2} \pm \sqrt{\frac{\lambda^2}{4}
-(\lambda^2-w^2)\left(c-\frac{1}{2}\right)^2} 
~,
\label{e3_15a}
\end{equation}

The form~(\ref{e3_15}) is hyperbolic for a weak interaction ($\lambda < w$) 
between subgraphs $A$ and $B$ and elliptic for strong coupling 
($\lambda > w$). 
The configurational averaging in Eq.~(\ref{e3_13a}) is straightforward and 
\begin{equation}
\langle g_1(\varepsilon)\rangle =\frac{N^{-1}|\varepsilon + r_0-w/2|}
{\sqrt{ (\lambda^2-w^2)(\lambda^2/4-(\varepsilon +r_0-w/2)^2)}}
\left[
\rho_c(c_1)+\rho_c(c_2)
\right]
 ~, 
\label{e3_16}
\end{equation}
where $c_{1,2}=1/2 \pm \left[(\lambda^2/4-(\varepsilon +r_0-w/2)^2)
/(\lambda^2-w^2) \right]^{1/2}$. 
The analysis of Eq.~(\ref{e3_16}) shows that, for any finite 
$N$, two  $\delta$-functions are 
broadened by disorder into two bands separated by a gap of width 
$W_g=\lambda$ for weak coupling and $W_g=w$ for strong coupling 
(see Fig.~\ref{f6}(a) and (b), respectively). 
The non-linearity of the map results in the singular behaviour 
of the ensemble-averaged DOS around $\varepsilon=-r_0+w/2\pm \lambda/2$ 
where $\langle g_1(\varepsilon)\rangle \propto (\varepsilon +  
r_0-w/2\mp \lambda/2)^{-1/2}$.

\begin{figure}[ht] %
\vskip2truecm 
\centerline{\includegraphics[width=13cm]{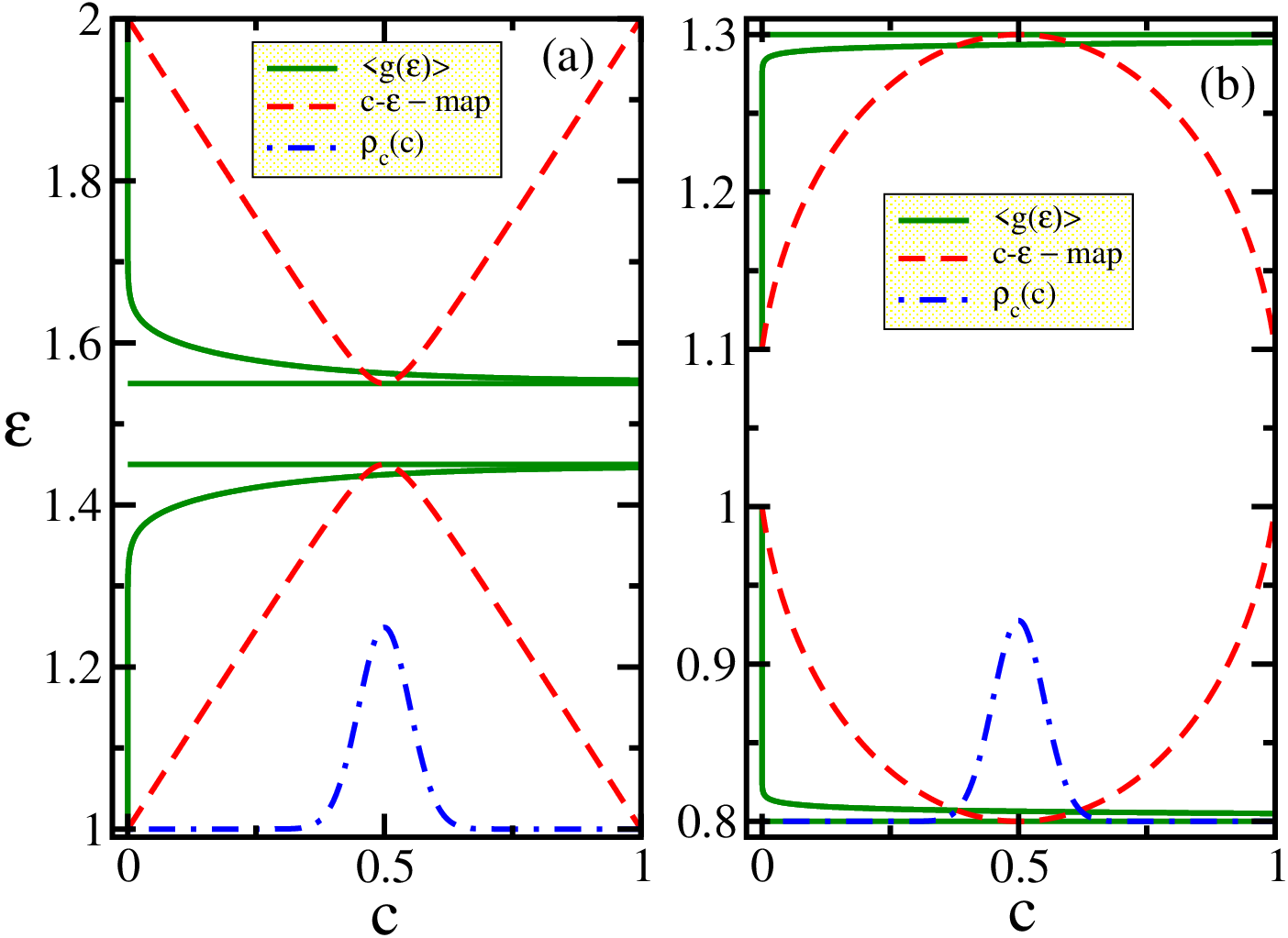}}  
\caption{(Color online) 
The $c-\varepsilon$ map (dashed line) for the symmetric binary FCG in the 
regime of weak (a) and strong (b) coupling. 
The ensemble-averaged DOS (scaled), $\langle g_1(\varepsilon)\rangle$, and the 
probability distribution function (scaled), $\rho_c(c)$, 
are shown by the solid and dot-dashed lines, respectively. 
The values of the parameters are: (a) $\epsilon_0=1$, $w=1$ and $\lambda=0.1$; 
(b) $\epsilon_0=1$, $w=0.1$ and $\lambda=0.5$. 
The probability $p=0.5$ and $N=100$ for both regimes. 
  }  
\label{f2} 
\end{figure}  

The ensemble-averaged spectrum for the binary FCG is bounded from the top 
by energy, $\varepsilon_{\text{max}}$, the knowledge of which is  
quite important from the applicational viewpoint (see below).  
The value of  $\varepsilon_{\text{max}}$ is easy to find from 
 Eqs.~(\ref{e3_15}), (\ref{e3_16}) and it is 
\begin{equation}
\varepsilon_{\text{max}}=-r_0 + w~~~\text{for}~~~\lambda<w
 ~, 
\label{e3_17}
\end{equation}
 and 
\begin{equation} 
\varepsilon_{\text{max}}=-r_0 + \frac{1}{2}(w+\lambda) 
~~~\text{for}~~~\lambda>w
 ~.  
\label{e3_18}
\end{equation}

Eqs.~(\ref{e3_17})-(\ref{e3_18}) give the upper boundaries for the  maximum 
eigenvalues for a particular realization of disorder, i.e. for a 
particular value of $c$. 
In the limit of large values of $N\to\infty$, the binomial  distribution 
$\rho_c(c)$ is of the Gaussian form characterized by the negligible width 
and thus it approaches the $\delta$-functional peak. 
Consequently, both energy bands collapse into two $\delta$-functions 
located at $\varepsilon = \varepsilon_{1,2}$ given by Eq.~(\ref{e3_15a}) and 
$\varepsilon_{\text{max}}\simeq \varepsilon_1$. 
Therefore, the critical value of parameter  $\eta=w/r_0$ can be found 
from the solution of equation, $\varepsilon_1=0$. 
In the limit of weak coupling  between the subgraphs, $\lambda < w$, 
the critical value $\eta_c^*$ depends 
on $c$ and lies in the interval, $ 1 \le \eta_c^* \le 2/(1+\lambda/w)$. 
The value of $\eta_c^*$ reaches the maximum, $\eta_c^*=2/(1+\lambda/w)$,
 for the homogeneous FCG, when $c=0$ 
(FCG contains only nodes of type $B$) or $c=1$ 
(FCG contains only nodes of type $A$). 
The minimum, $\eta_c^*=1$, is attained  for 
equal concentrations of nodes $A$ and $B$, i.e. when $c=0.5$. 
In the regime of strong coupling between subgraphs, $\lambda > w$, 
the situation changes to the opposite one so that 
$ 2/(1+\lambda/w)\le \eta_c^* \le 1$ and the lowest value of 
$\eta_c^*$ corresponds to $c=0.5$. 

Biologically, this means that  mixed populations containing two species 
are most vulnerable to epidemics for equal concentrations of species 
if these species strongly interact with each other 
(in terms of transferring disease) 
and if the species do not interact strongly enough then the 
mixture of species enhances the resistance of the population. 

The above estimates hold for the populations of organisms which are able 
to communicate (transfer a disease) to all other members of the population, 
i.e. for populations having the communication network with the topology 
of the FCG. The other assumption is that the transmission rate between 
two individuals 
is inversely proportional to the number of individuals in the population. 
This is a plausible assumption for migrating individuals when the 
probability to establish a contact can be proportional to $N^{-1}$.

\subsection{Lloyd's model} 
\label{s2_2}
It is known that 
the configurational averaging of the spectral density 
can be performed analytically for the network of any topology, and thus 
for the FCG as well, 
if the diagonal elements of the Hamiltonian matrix are distributed according 
to  the Cauchy distribution, 
\begin{equation}
\rho_{\epsilon_i}(\epsilon_i)=\rho_{\epsilon}(\epsilon_i)=
\frac{\delta}{\pi}\,\frac{1}{\delta^2+(\epsilon_i-\epsilon_0)^2}
~,
\label{e2_2_1}
\end{equation}
where $\delta$ is the width of the distribution, and all the relevant 
node-node interactions are not random, $V_{ij}=V_0$ 
(Lloyd's model \cite{Lloyd_69}).  

The diagonal elements of the configurationally-averaged resolvent operator, 
$\langle G_{ii}(\varepsilon)\rangle$ can be expressed 
via the resolvent elements for the ideal FCG, $G_{ii}^{(0)}$, with the 
argument shifted to the upper half of the complex plane, 
$\langle G_{ii}(\varepsilon)\rangle=
G_{ii}^{(0)}(\varepsilon + \text{i}\delta)$. 
Therefore the ensemble-averaged spectral density for Lloyd's model 
defined on the FCG has the following form: 
\begin{equation}
\langle g(\varepsilon)\rangle = \left(1-\frac{1}{N}\right)
\rho_{\epsilon}(\varepsilon -V_0-\gamma(N-1)V_0)
+ 
\frac{1}{N}\rho_{\epsilon}(\varepsilon-(\gamma-1)(N-1)V_0) 
~,
\label{e2_2_2}
\end{equation}
i.e. the DOS is obtained from that for the ideal FCG by broadening of 
both $\delta$-functions in Eq.~(\ref{e2_1_2}) into two Lorentzian peaks. 
Formally, the FCG with on-site energies distributed according to the 
Cauchy distribution is equivalent to the ideal FCG with nodes characterized 
by the complex on-site energies (see also \cite{Neut_95}). 

In contrast to the binary FCG, 
the widths of both of the Lorentzian peaks in Eq.~(\ref{e2_2_2}) do not depend 
on $N$ and for any value of $\delta$ 
it is possible to find such a value of $N$ starting from which the 
positive eigenvalues appear in the spectrum. 
This means that Lloyd's model on the FCG is not resistant to the invasion 
of epidemics at least in the dilute regime. 
It is a consequence of the special form of the distribution of the 
recovery rates given by Eq.~(\ref{e2_2_1}) with not existing high moments.

\section{Polynomial interactions} 
\label{s3}

We have considered above several simple examples of the FCG for which 
the spectrum can be found exactly and its configurational averaging can be 
performed analytically. 
A natural question arises about the possibility of a similar analysis 
in the case of more general disorder. 
Below, we demonstrate that indeed in the case of the polynomial 
node-node interactions such an analysis is possible.  

\subsection{General Solution} 
\label{s3_1}

In the case of the polynomial node-node interactions for the FCG, the 
matrix elements of the resolvent operator in the site basis can 
be found exactly  
and the spectral density for a particular realization of disorder 
is given by the following formula (see Appendix~\ref{app_a}): 
\begin{equation}
g(\varepsilon)=g_0(\varepsilon) + \delta g(\varepsilon)
= \frac{1}{N} \sum_i^N \delta(\varepsilon - \widetilde{\epsilon}_i) 
- \frac{1}{\pi N}\text{Im}\,
\frac{\text{d}}{\text{d}\varepsilon}
 \left[ 
\ln \left(\text{det}\,{\bf D}(\varepsilon)\right)
\right]
~.
\label{e2_2}
\end{equation}
Here the renormalized bare energies, $\widetilde{\epsilon}_i$, are given 
by Eq.~(\ref{app_a_2}) and the spectral determinant, 
$\text{det}\,{\bf D}(\varepsilon)$, satisfies  Eq.~(\ref{app_a_7}). 

Assuming for simplicity that all $\epsilon_i$ and $\phi_i$ are 
identically distributed according to the probability distribution 
functions $\rho_{\epsilon_i}(\epsilon_i)\equiv \rho_{\epsilon}(\epsilon_i) $ 
and  $\rho_{\phi_i}(\phi_i)\equiv \rho_{\phi}(\phi_i) $, respectively, 
the spectral density $g(\varepsilon)$ can be configurationally averaged 
according to Eq.~(\ref{e1_3}), 
\begin{eqnarray}
\langle g(\varepsilon)\rangle 
&=& 
\langle g_0(\varepsilon)\rangle + \langle \delta g(\varepsilon)\rangle  
\nonumber \\
&=&
\int\limits_{-\infty}^{\infty} 
\rho_{\phi}(\phi_i) \rho_{\epsilon}
(\epsilon - \widetilde{V}_{ii})\,\text{d}\phi_i
- \frac{1}{\pi N}\,
\frac{\text{d}}{\text{d}\varepsilon}
 \left\langle
\text{arg}\,\left(\text{det}\,{\bf D}(\varepsilon)\right) 
\right\rangle
~,
\label{e2_3}
\end{eqnarray}
where 
\begin{equation}
\widetilde{V}_{ii} =  
\gamma \bm{\psi}_i^{T}\sum_j^N \bm{\varphi}_j+
(1-\gamma)\bm{\psi}_i^{T}\bm{\varphi}_i
\simeq 
\gamma N \bm{\psi}_i^{T}\overline{\bm{\varphi}}+
(1-\gamma)\bm{\psi}_i^{T}\bm{\varphi}_i
~,  
\label{e2_3a}
\end{equation}
with $\bm{\psi}_i^T=\bm{\varphi}_i^T \bm{\alpha}$ and 
$\overline{\bm{\varphi}} = \int_{-\infty}^{\infty} 
\bm{\varphi}_i \rho_{\phi}(\phi_i)\, 
\text{d}\phi_i  $ so that 
$\widetilde{V}_{ii} $ depends on the characteristics of node $i$ only  
 (see below for the justification of this approximation). 
The ensemble-averaged spectral density, $\langle g(\varepsilon) \rangle$, 
has two contributions. 
The first contribution, $\langle g_0(\varepsilon) \rangle$, 
is given by the convolution of two probability distributions, one of them, 
$\rho_{\epsilon}$, is shifted along the energy axis due to the node-node 
interactions. 
If both distributions $\rho_{\epsilon}$ and $\rho_{\phi}$ are 
band-shaped 
(e.g. normal, box or $\delta$-functional 
distributions) 
 then the function  $\langle g_0(\varepsilon) \rangle$ also has 
the shape of a band  of typical width $\Delta$  
defined via the widths of the distributions 
 $\rho_{\epsilon}$ and $\rho_{\phi}$.

The other contribution to the spectral density comes from 
$\langle \delta g(\varepsilon)\rangle $ and its magnitude 
is negligible in the main band region 
(due to the factor $N^{-1}$ in Eq.~(\ref{e2_3})) and finite outside the 
band with $\langle \delta g(\varepsilon)\rangle $ being in the form 
of several peaks (see below). 
The functional form of $\langle \delta g(\varepsilon)\rangle $ 
depends on the properties of the spectral determinant
$\text{det}\,{\bf D}$. 
It follows from Eq.~(\ref{app_a_7}), that the spectral determinant is 
a random value which depends only on macroscopic sums, 
$a_k = \sum_{i}^N a_{ki} = \sum_{i}^N \phi_i^{k}
(\varepsilon-\tilde{\epsilon}_i)^{-1}$, i.e. 
$\text{det}\,{\bf D}= D(\varepsilon; a_0,a_1,\ldots,a_{2n-2})$. 
According to the central limit theorem, the values of $a_k$ are distributed 
around the mean value $\overline{a_k}\simeq N \overline{a_{ki}}$ in the peak 
region of width $\delta a_k \simeq \left(N \text{Var}[a_{ki}]\right)^{1/2}$, 
i.e. the relative peak width of this distribution approaches zero 
in the thermodynamic limit 
($N\to \infty$), $\delta a_k/\overline{a_k} \propto N^{-1/2} \to 0$. 
Bearing this in mind we can perform configurational averaging 
of the phase of the spectral determinant in Eq.~(\ref{e2_3}) 
approximately (assuming, in fact, that the spectral density is 
a self-averaging quantity), 
\begin{equation}
\left\langle
\text{arg}\,\left(\text{det}\,{\bf D}
(\varepsilon; a_0,a_1,\ldots,a_{2n-2})\right) 
\right\rangle \simeq 
\text{arg}\,\left(\text{det}\,{\bf D}
(\varepsilon; \overline{a_0},\overline{a_1},
\ldots,\overline{a_{2n-2}})\right)
~, 
\label{e2_4}
\end{equation}
where 
\begin{eqnarray}
\overline{a_k(\varepsilon)} &=& 
N \iint\limits_{-\infty}^{\infty} \frac{\phi_i^{k}}{\varepsilon - \epsilon_i-
\widetilde{V}_{ii}(\phi_i)}\, 
\rho_{\epsilon}(\epsilon_i)\rho_{\phi}(\phi_i)\,\text{d}\epsilon_i\,
\text{d}\phi_i  
\nonumber \\
&=& N \int\limits_{-\infty}^{\infty} \phi_i^{k} R(\varepsilon - 
\widetilde{V}_{ii}(\phi_i))\,\text{d}\phi_i 
+\text{i} N 
\int\limits_{-\infty}^{\infty} \phi_i^{k} I(\varepsilon - 
\widetilde{V}_{ii}(\phi_i))\,\text{d}\phi_i 
~, 
\label{e2_5}
\end{eqnarray}
with 
\begin{equation}
R(z)=\diagup\hskip-15pt\int\limits_{-\infty}^{\infty}\, 
\frac{\rho_{\epsilon}(\epsilon_i)}
{z-\epsilon_i}\, \text{d}\epsilon_i~~~~\text{and}~~~~
I(z)= -\pi  \rho_{\epsilon}(z) 
~.   
\label{e2_6}
\end{equation}
Eq.~(\ref{e2_4}) becomes exact for an infinite number of nodes. 
Note that the same arguments were used in the derivation of 
 Eq.~(\ref{e2_3a}) for 
replacing $ \sum_j^N \bm{\varphi}_j$ with $N \overline{\bm{\varphi}}$ 
resulting in $\widetilde{V}_{ii}$ 
being a function of $\phi_i$ only but not other 
$\phi_j$ for $j\ne i$. 
 
Expressing the phase, 
$\text{arg}\,\left(\text{det}\,{\bf D}
(\varepsilon; \overline{a_0},\overline{a_1},
\ldots,\overline{a_{2n-2}})\right)$,  via real and imaginary parts 
of the spectral determinant, 
$D=\widetilde{D}+\text{i}\widetilde{\widetilde{D}}$, 
and differentiating it with respect to energy according to 
Eq.~(\ref{e2_3}) we arrive at  
the following final expression for 
 $\langle g(\varepsilon)\rangle$, 
\begin{equation}
\langle  g(\varepsilon)\rangle = \langle  g_0(\varepsilon)\rangle+
\langle \delta g(\varepsilon)\rangle 
\simeq 
\int\limits_{-\infty}^{\infty} \rho_{\phi}(\phi_i) \rho_{\epsilon}
(\epsilon - \widetilde{V}_{ii})\,\text{d}\phi_i 
-\frac{1}{\pi N}\, \frac{\widetilde{D}\left.\widetilde{\widetilde{D}}\right.'- 
\left.\widetilde{D}\right.'\widetilde{\widetilde{D}}}
{\left.{\widetilde{D}}\right.^2 + \left.\widetilde{\widetilde{D}}\right.^2}
~, 
\label{e2_7}
\end{equation}
where prime means differentiation with respect to $\varepsilon$. 
This is the main result of the paper. 
Eq.~(\ref{e2_7}) allows the spectral density of the FCG with polynomial 
interactions to be calculated for rather general distributions of the 
bare energies, $\epsilon_i$, and interaction characteristics, $\phi_i$. 
The functional form of the spectral determinant $D(\varepsilon)$ depends 
on a concrete formulation of the problem but it is irrelevant for the 
main band shape and can influence only the positions of discrete levels split 
from the main band. 
An alternative derivation of Eq.~(\ref{e2_7}) by means of direct integration 
of the left-hand side of Eq.~(\ref{e2_4}) (and thus demonstrating the 
self-averaging property for the spectral density) is presented in 
Appendix~\ref{app_b} for multiplicative interactions in the electronic case. 

In the band region, where the  contribution from 
 $\langle g_0(\varepsilon)\rangle $ is significant, both functions 
$\widetilde{D}(\varepsilon)$ and 
$\widetilde{\widetilde{D}}(\varepsilon)$ are typically of the same order 
of magnitude, $\widetilde{D} \sim \widetilde{\widetilde{D}}$  and 
 $\left.\widetilde{D}\right.' \sim \left.\widetilde{\widetilde{D}}\right.' 
\sim \widetilde{D}/\Delta $ (if $n\ll N$),  
so that the contribution from $\langle \delta g(\varepsilon)\rangle $ 
to the total ensemble-averaged spectral density is 
macroscopically small, 
$\langle \delta g(\varepsilon)\rangle \sim (N\Delta)^{-1}$. 
This is not surprising and is a consequence of the particular form 
of the node-node interactions given by Eq.~(\ref{e1_2}) forcing the 
majority of the eigenvalues, $\varepsilon_i$, of the Hamiltonian~(\ref{e1_1})  
(the roots of the spectral determinant, $D(\varepsilon_i)=0$) 
to be bound between the consequent renormalized bare energies, 
i.e. 
$\varepsilon_i \in (\widetilde{\epsilon}_j,\widetilde{\epsilon}_{j+1})$. 
This property is similar to the well-known phenomenon of the 
spectral reconstruction caused by the interactions of one level 
(e.g. associated 
with the defect) with a continuum of levels (band) 
(see e.g. Refs.~
\cite{Maradudin_71,Lehmann_73,Bohr_89:book,Klinger_93,Bogomolny_01} 
and the discussion  in Sec.~\ref{s2_3_1}).

Outside the main band region, where $\widetilde{\widetilde{D}}\to 0$, the 
function $\langle \delta g(\varepsilon)\rangle $ contributes  
in the form of the Gaussian peaks 
(see the explanation below),  
$\langle \delta g(\varepsilon)\rangle = \sum_{i=1}^{m} 
N(E[\varepsilon_{*i}],\text{Var}[\varepsilon_{*i}])$, of width 
$\delta\varepsilon_{*i}=\sqrt{\text{Var}[\varepsilon_{*i}]}$ and 
centred around the expectation value, $E[\varepsilon_{*i}]$ 
(the function $N(E[x],\text{Var}[x])$ 
stands for the normal distribution of $x$ with 
the expectation value $E[x]$ and variance $\text{Var}[x]$). 
The peak locations coincide with the roots of the real part of the 
spectral determinant, $\widetilde{D}(E[\varepsilon_{*i}])=0$, and the number 
of roots, $m$, cannot exceed the order of the spectral determinant, 
i.e. the order of interactions, $m\le n$. 
Indeed, for energies far away from the band, ${\cal E}/\Delta \gg 1 $ where 
${\cal E}= |\varepsilon -\overline{\epsilon} -
\overline{\widetilde{V}_{jj}}|$ (where $\overline{\widetilde{V}_{jj}} = 
\int \widetilde{V}_{jj}\rho_{\phi}(\phi_j)\text{d}\phi_i$ and 
$\overline{\epsilon} = \int \epsilon_i \rho_{\epsilon}(\epsilon_i)
\text{d}\epsilon_i$), 
we can estimate the value of 
$\overline{a}_k \sim N\overline{\phi^{k}}/{\cal E} $, so that the spectral 
determinant is the $n$-th order polynomial of $(N/{\cal E})$ with constant finite 
coefficients (under the assumption that the interaction coefficients 
$\alpha_{ij}$ do not depend on $N$).  
It can have maximum $n$ of $N$-independent roots, so that 
$|\varepsilon_{*i}-\overline{\epsilon} -\overline{\widetilde{V}_{jj}}| 
\propto N $. 
This means that the Gaussian peaks are mainly macroscopically 
separated from the band. 

Therefore, the interactions of polynomial type~(\ref{e1_2}) change 
the bare spectrum $\epsilon_i$ of the Hamiltonian~(\ref{e1_1}) in the 
following manner: (i) the bare band is shifted and deformed and 
(ii) several isolated levels macroscopically separated from 
the band are formed. 
Such a picture will be supported below by detailed analysis of some simple 
cases of the low-order interactions for $n=2$.  
    
\subsection{\label{s2_3_1} 
Multiplicative interactions for the electronic problem} 

We start the analysis with the simplest case of multiplicative (separable) 
node-node interactions, when the second-order interaction matrix 
contains only one non-zero element, $\alpha_{22}$, so that 
\begin{equation}
V_{ij}=\left(1~~\phi_i\right)
\left(
\begin{array}{cc}
0 ~&~0 \\
0 ~&~\alpha_{22}
\end{array}
\right)
\left(
\begin{array}{c}
1 \\
\phi_j 
\end{array}
\right)
= \alpha_{22}\phi_i \phi_j
~. 
\label{e2_3_1_1}
\end{equation}
The spectral determinant for a particular realization of disorder is
\begin{equation}
D(\varepsilon) = 1+\sum_i^N \frac{\alpha_{22}\phi_i^2}
{\varepsilon - \tilde\epsilon_i}
~, 
\label{e2_3_1_2} 
\end{equation}
with 
\begin{equation}
\tilde\epsilon_i = \epsilon_i +\widetilde{V}_{ii} 
 \simeq \epsilon_i +\gamma\alpha_{22}N \phi_i \overline{\phi} 
+(1-\gamma)\alpha_{22}\phi_i^2
~.  
\label{e2_3_1_3} 
\end{equation}
We consider below the electronic problem ($\gamma=0$) mainly but 
the general results for arbitrary values of $\gamma$ will be presented 
when possible.  

The roots of the spectral determinant, $\varepsilon_i$, 
are bound between the renormalized bare energies, 
$\varepsilon_i \in (\tilde{\epsilon}_{i-1},\tilde{\epsilon}_{i})$ 
(for $\alpha_{22}>0$) which is obvious from the functional form 
of $D(\varepsilon)$ given by Eq.~(\ref{e2_3_1_2}). 
Therefore the eigenvalues of the Hamiltonian 
are expected to be very close to the renormalized bare energies, 
$\varepsilon_i = \tilde{\epsilon_i}+ O(\Delta/N)$, and the changes  
in the spectral density within the band region for $\tilde{\epsilon_i}$ 
due to interactions~(\ref{e2_3_1_1}) should be negligible ($\propto N^{-1}$). 
This property  is very similar 
to Rayleigh's theorem in the theory of vibrations in disordered 
systems \cite{Maradudin_71}. 

The ensemble-averaged spectral density is 
given by Eq.~(\ref{e2_7}) 
with
\begin{equation}
\widetilde{D}(\varepsilon;\overline{a_2}) = 
1+ \text{Re}\,\overline{a_2(\varepsilon)} = 
1 + N  \int\limits_{-\infty}^{\infty} \alpha_{22} 
\phi_i^{2} R(\varepsilon - \widetilde{V}_{ii})\,\text{d}\phi_i 
~,  
\label{e2_3_1_4} 
\end{equation}
and 
\begin{equation}
\widetilde{\widetilde{D}}(\varepsilon;\overline{a_2}) = 
 \text{Im}\,\overline{a_2(\varepsilon)} = 
N  \int\limits_{-\infty}^{\infty} \alpha_{22} 
\phi_i^{2} I(\varepsilon - \widetilde{V}_{ii})\,\text{d}\phi_i 
~,    
\label{e2_3_1_5} 
\end{equation}
where the functions $R(z)$ and $I(z)$ are defined by 
Eq.~(\ref{e2_6}).  
The properties of the function $\langle g(\varepsilon)\rangle $ have 
already been discussed above. 
In particular, in the main-band region the contribution from 
 $\langle \delta g(\varepsilon)\rangle $ is negligible 
($\propto (N\Delta)^{-1}$) while outside the band it can exhibit  
Gaussian peaks.  
For multiplicative interactions, there is only one Gaussian peak 
outside the band, $\langle g(\varepsilon)\rangle \simeq 
\langle \delta g(\varepsilon)\rangle 
\simeq N(\varepsilon_*,\sigma^2)$, with the peak position being the 
solution of the following equation $\widetilde{D}(\varepsilon_*)=0$, i.e.
\begin{equation}
1 + N  \int\limits_{-\infty}^{\infty} \alpha_{22} 
\phi_i^{2} R(\varepsilon_* - \widetilde{V}_{ii})\,\text{d}\phi_i = 0 
~.   
\label{e2_3_1_6} 
\end{equation}
This equation can be solved approximately assuming that the level 
$\varepsilon_*$  
is split from the band far enough in comparison with the band width, 
i.e. $\left|\varepsilon_*-\overline{\epsilon} - 
\overline{\widetilde{V}_{ii}}\right|/\Delta \gg 1$, 
\begin{equation}
\varepsilon_* \simeq \overline{\epsilon} 
- N \alpha_{22} \overline{\phi^{2}}
~.  
\label{e2_3_1_7} 
\end{equation}
Indeed, we see from Eq.~(\ref{e2_3_1_7}) that the distance between the 
isolated level and the main band is macroscopically large, 
$\propto N$ (if the 
coefficient $\alpha_{22}$ does not depend on $N$), which is consistent 
with the result for the ideal FCG (see Eq.~(\ref{e2_1_2})) showing the 
similar structure of the spectrum with the $(N-1)$-degenerate level playing 
the role of the main band. 

In order to estimate the width of the Gaussian peak, we consider a particular 
$j$-th realization of disorder for which the random value of 
$\varepsilon_{j*}$ can be estimated in a similar manner, 
$ \varepsilon_{j*}\simeq N^{-1}\sum_i^N \epsilon_i - 
\alpha_{22} \sum_i^N \phi_i^2$. 
From this expression according to the central limit theorem,  we conclude  
 that the values of 
$\varepsilon_{j*}$ are normally distributed with the expectation 
value given by Eq.~(\ref{e2_3_1_7}) and with variance, 
\begin{equation}
\sigma^2 \simeq \frac{\text{Var}[\epsilon_i]}{N} +
N \alpha_{22}^2 \text{Var}[\phi_i^2]
~.  
\label{e2_3_1_8} 
\end{equation}
This expression for the variance coincides with that given by 
Eq.~(\ref{app_b_22}) more rigorously derived in Appendix~\ref{app_b}. 
It follows from Eq.~(\ref{e2_3_1_8}) that the peak width, $\sigma$, 
depends on $N$ and 
it increases with $N$, $\sigma \propto \sqrt{N}$, if the interaction 
coefficient $\alpha_{22}$ does not depend on $N$. 
However, if $\alpha_{22} \propto 1/N$, the peak width decreases with $N$, 
$\sigma \propto 1/\sqrt{N}$, and the peak collapses to a $\delta$-function 
in the thermodynamic limit. 

All the results presented above for the multiplicative interactions are  
supported in Appendix~\ref{app_b} 
by an alternative derivation of Eqs.~(\ref{e2_7}), 
(\ref{e2_3_1_4})-(\ref{e2_3_1_5}) for the ensemble-averaged spectral 
density using direct approximate integration of Eq.~(\ref{e1_3}). 

\begin{figure}[ht]  
\vskip2truecm
\centerline{\includegraphics[width=13cm]{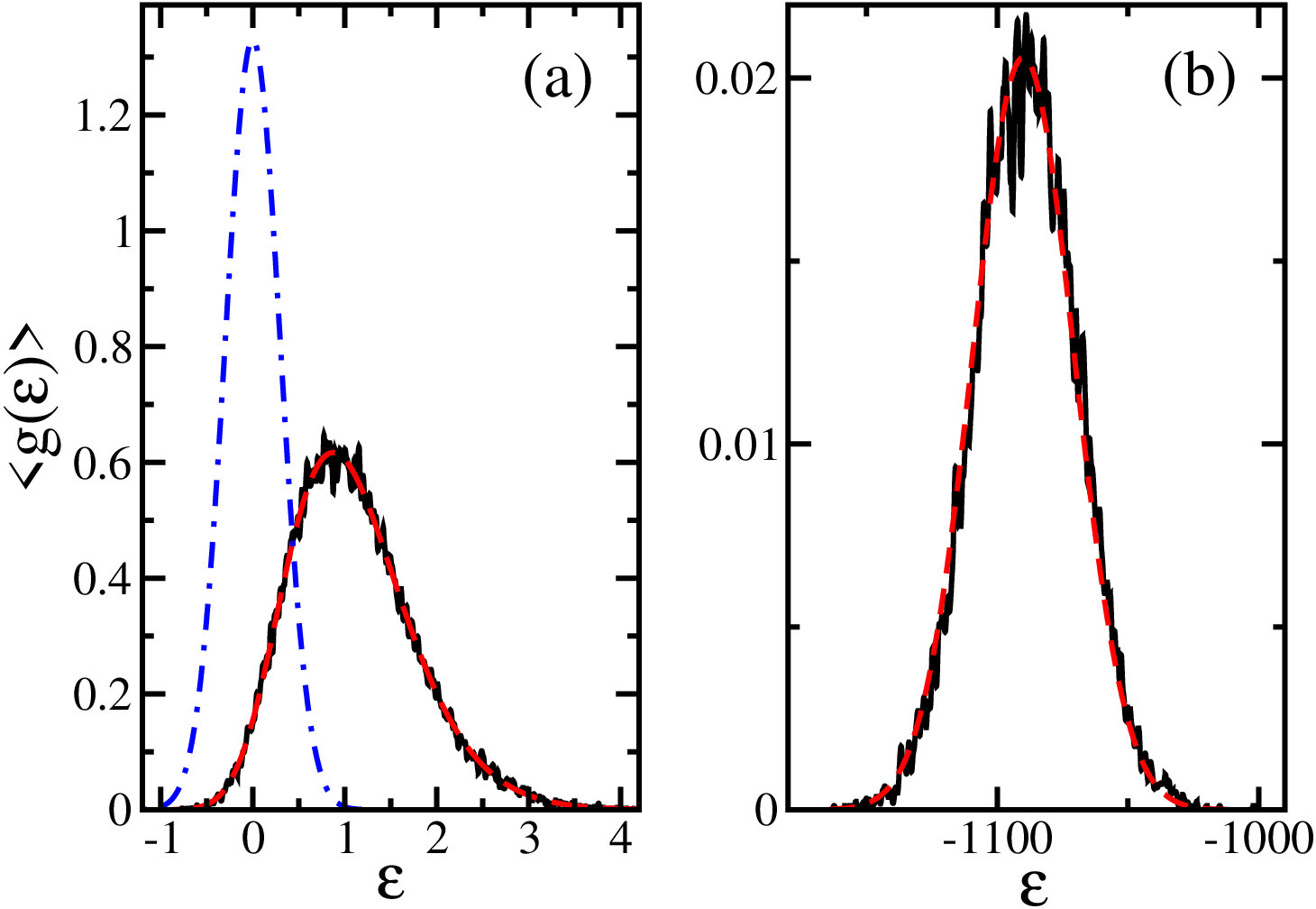}}  
\caption{(Color online)
The ensemble-averaged DOS (the main band, 
$\langle g(\varepsilon) \rangle \simeq \langle g_0(\varepsilon) \rangle$, 
in (a) 
and separate level,  
$\langle g(\varepsilon) \rangle \simeq \langle \delta g(\varepsilon) \rangle$, 
outside the main band  
in (b)) for electron problem 
($\gamma=0$) defined on the FCG with multiplicative 
interactions. 
Both the on-site energies $\epsilon_i$ and 
interaction parameters $\phi_i$ are 
normally distributed according to  $\rho_\epsilon(\epsilon_i)=
N(0,0.09)$ (the dot-dashed line in Fig.~\ref{f1}(a)) 
and  $\rho_\phi(\phi_i)=N(1,0.09)$; $\alpha_{22}=1$. 
The dashed lines in (a) and (b) are plotted according to the first term in 
Eq.~(\ref{e2_7}) and  Eqs.~(\ref{app_b_21})-(\ref{app_b_22}), respectively. 
The solid lines represent the data obtained by direct diagonalization 
of the Hamiltonian  for $N=1000$ nodes and averaging over 
$10^4$ configurations. 
The DOS in  Fig.~\ref{f3}(b) is scaled by factor $N$.   
  }  
\label{f3} 
\end{figure}

Numerical analysis confirms all the conclusions made above. 
Fig.~\ref{f3} demonstrates both the main band (a) and the Gaussian peak (b) 
for the case of multiplicative interactions for the electronic problem 
($\gamma=0$) defined on the FCG. 
It is clearly seen how the interactions shift the Gaussian bare 
band (dot-dashed curve in Fig.~\ref{f3}(a)) and change its shape. 
The isolated level is normally distributed (see Fig.~\ref{f3}(b)) and 
macroscopically shifted down along the energy axis. 
For all cases, the results of direct diagonalization are practically  
indistinguishable 
from those obtained in accord with analytical expressions.   
 
\begin{figure}[ht] %
\vskip1truecm
\centerline{\includegraphics[width=10cm]{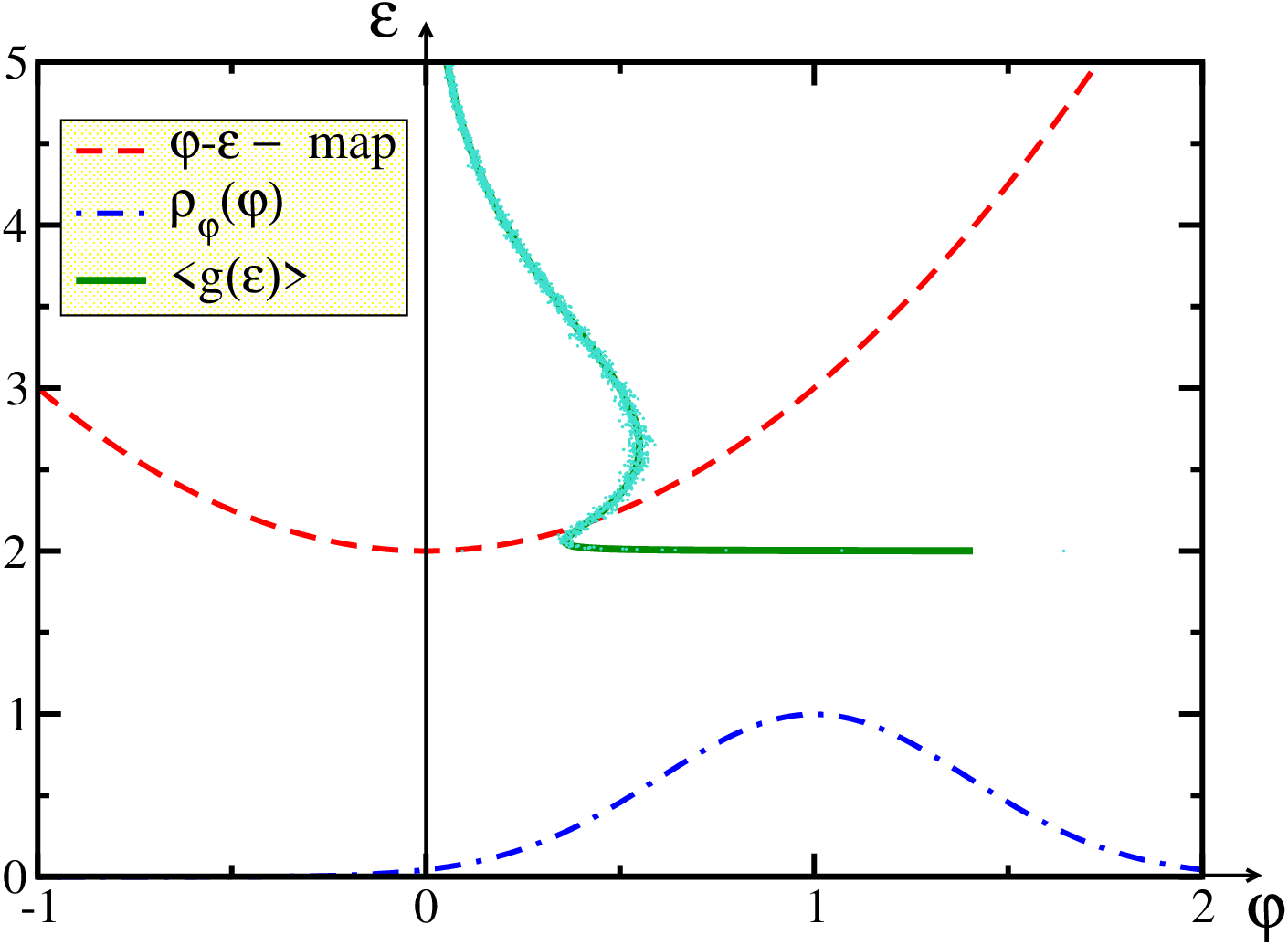}}  
\caption{(Color online)
The $\phi-\varepsilon$ parabolic map (dashed line) 
for the electron problem ($\gamma=0$) defined on the FCG 
with multiplicative interactions ($\epsilon_0=2$, $\alpha_{22}=1$).  
The interaction parameters are normally distributed 
according to $\rho_\phi(\phi_i)=N(1,0.16)$.
The ensemble-averaged DOS (scaled), 
$\langle g(\varepsilon)\rangle \simeq \langle g_0(\varepsilon)\rangle$, 
and the 
probability distribution function (scaled), $\rho_\phi(\phi_i)$, 
are shown by the solid 
(Eq.~(\ref{e2_3_1_9})) and dot-dashed lines, respectively. 
The dots scattered around the solid line were obtained by 
direct diagonalization of the Hamiltonian ($N=3000$) and by averaging 
over 300 realizations of disorder. 
  }  
\label{f4} 
\end{figure}

\begin{figure}[ht] %
\centerline{\includegraphics[width=10cm]{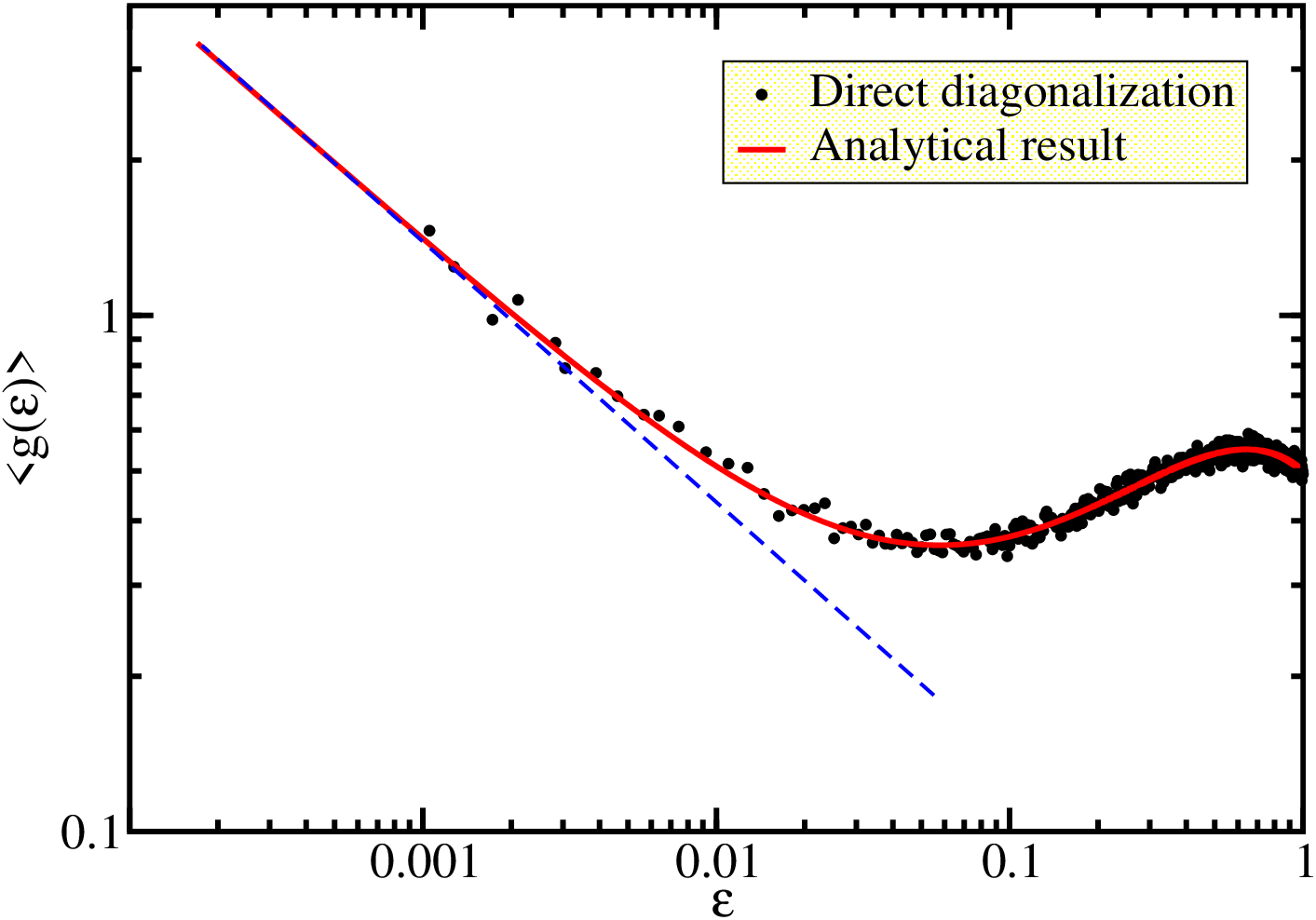}}  
\caption{(Color online) 
The double-log plot of 
$\langle g(\varepsilon)\rangle \simeq \langle g_0(\varepsilon)\rangle$ 
shown in Fig.~\ref{f4} around the singularity. 
The solid line was obtained using Eq.~(\ref{e2_3_1_10}) and the dots represent 
the results of direct diagonalization. 
The dashed line, $\propto \varepsilon^{-1/2}$, is used as an eye guide. 
  }  
\label{f5} 
\end{figure}

The evaluation of the integral in the expression for 
$\langle g_0(\varepsilon) \rangle$ (see Eq.~(\ref{e2_7})) has been 
performed numerically in the above example illustrated in Fig.~\ref{f3}. 
Such an integration  becomes trivial if one of the 
probability distribution functions is a $\delta$-function. 
For example, if the on-site energies are randomly distributed while 
the interactions are all the same, 
$\rho_{\phi}(\phi_i)=\delta(\phi_i-\phi_0)$, the configurationally 
averaged spectral density coincides with the shifted 
distribution of the on-site energies,  
$ \langle g_0(\varepsilon)\rangle = 
\rho_{\epsilon}(\varepsilon -\alpha_{22}\phi_0^2)$ 
(due to the linear map between $\epsilon_i$ and $\varepsilon$, i.e. 
$\varepsilon = \tilde{\epsilon}_i=\epsilon_i+\alpha_{22}\phi_0^2$ which 
follows from Eq.~(\ref{e2_3_1_3})). 
On the other hand, the on-site energies can be all the same, 
$\rho_{\epsilon}(\epsilon_i)=\delta(\epsilon_i-\epsilon_0)$, but 
the interactions are randomly distributed. 
In this case, due to the quadratic map between $\phi_i$ and $\varepsilon$, 
i.e. $\varepsilon = \tilde{\epsilon}_i=\epsilon_0+\alpha_{22}\phi_i^2$ 
(see Eq.~(\ref{e2_3_1_3})), there are two contributions to the 
ensemble-averaged DOS from different branches of this quadratic map  
(see Fig.~\ref{f4}), 
\begin{equation}
\langle g_0(\varepsilon)\rangle = 
\frac{1}{2\sqrt{\alpha_{22}(\varepsilon-\epsilon_0)}}
\left[ 
\rho_{\phi}\left(\sqrt{\frac{\varepsilon-\epsilon_0}{\alpha_{22}}} \right)+ 
\rho_{\phi}\left(-\sqrt{\frac{\varepsilon-\epsilon_0}{\alpha_{22}}} \right)
\right]
~.
\label{e2_3_1_9}
\end{equation}
with $\varepsilon$ obeying the following inequality, 
$\alpha_{22}(\varepsilon-\epsilon_0)>0$. 
The ensemble-averaged DOS exhibits a singular behaviour around the 
boundary of the spectrum at 
$\varepsilon \simeq \epsilon_0$
(see Fig.~\ref{f5}), 
\begin{equation}
\langle g_0(\varepsilon)\rangle \simeq 
\frac{\rho_{\phi}(0)}{\sqrt{\alpha_{22}(\varepsilon-\epsilon_0)}}
 ~,
\label{e2_3_1_10}
\end{equation}
if $\rho_{\phi}(0)$ is a finite non-zero value. 
From this example, we see that the bare spectral density can display  
quite drastic changes in shape depending on the type and probability 
distribution of the interaction parameters.


In terms of epidemiological applications, the above analysis can be used 
for an estimate of the critical parameter $\eta_c$. 
Indeed, 
the low-bound estimate for $\eta_c$ can be easily obtained for the FCG with 
polynomial interactions by finding the position of the 
largest isolated eigenvalue, $\varepsilon_*^{\text{max}}$, and solving 
the equation, $\varepsilon_*^{\text{max}}(\eta_c^*)=0$ with  
$\varepsilon_*^{\text{max}}$ being the largest root of 
the real part of the spectral determinant, 
$\widetilde{D}(\varepsilon_*^{\text{max}})=0$. 
An important and  general conclusion which follows from the analysis 
presented above is that the isolated normally distributed roots of the 
real part of the spectral determinant scale linearly with $N$, 
i.e. $\varepsilon_*^{\text{max}} \propto N$ if the interaction 
coefficients, $\alpha_{ij}$, and the interaction parameters, $\phi_i$, 
 and thus the transmission rates, $W_{ij},$ do not depend on $N$. 
This means that equation   $\varepsilon_*^{\text{max}}(\eta_c^*)=0$ 
does not have the solution independent of $N$ 
and the system does not exhibit the phase 
transition in the thermodynamic limit  (it is always in the active state). 
In the opposite case, when the transmission rates are inversely proportional 
to $N$, the maximum eigenvalue does not depend on $N$ and the transition 
exists at least for the Hamiltonian in the dilute regime. 
The concrete value of the critical parameter $\eta_c^*$ depends on the  
particular 
type of the interactions and the probability distributions for recovery 
and transmission rates.

\subsection{\label{s2_3_2} 
Multiplicative interactions for the vibrational problem} 

\begin{figure}[ht] %
\vskip1truecm
\centerline{\includegraphics[width=10cm]{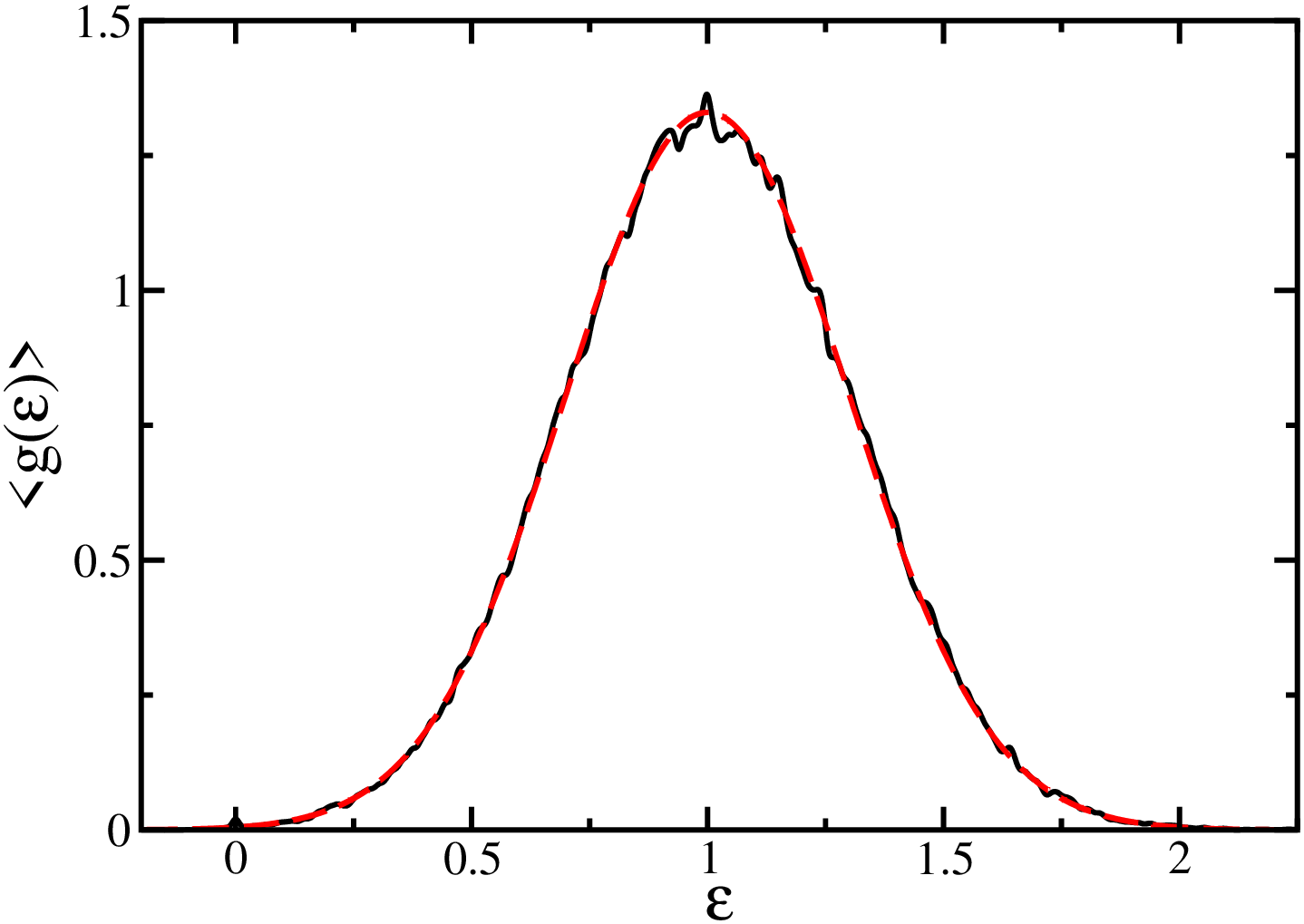}}  
\caption{(Color online) 
The ensemble-averaged DOS for the vibrational problem 
($\gamma=1$) defined on the FCG with multiplicative 
interactions. 
The random interaction parameters $\phi_i$ are 
normally distributed according to $\rho_\phi(\phi_i)=
N(1,0.09)$; $\beta=1$. 
The dashed line is plotted according to Eq.~(\ref{e2_3_2_2}) and the 
data for the solid line are obtained by direct diagonalization 
of the Hamiltonian  for $N=3000$ nodes and averaging over 
50 configurations. 
  }  
\label{f6} 
\end{figure}

For vibrational problem defined on the FCG ($\gamma=1$ and $\epsilon_i=0$) 
with multiplicative interactions, 
the ensemble-averaged DOS is given by Eq.~(\ref{e2_7}) with 
$\rho_{\epsilon_i}(\epsilon_i)=\delta(\epsilon_i)$ and 
\begin{equation}
\tilde\epsilon_i = \alpha_{22}N\phi_i\overline{\phi}\equiv
\beta \phi_i\overline{\phi}
~, 
\label{e2_3_2_1}
\end{equation}
where $\beta=\alpha_{22}N$. 
Eq.~(\ref{e2_3_2_1}) reveals the linear map between  
$\phi_i$ and $\varepsilon$, 
i.e. $\varepsilon = \tilde{\epsilon}_i=\beta \phi_i\overline{\phi}$, 
and thus for the main band,  
\begin{equation}
\langle g_0(\varepsilon)\rangle = \frac{1}{|\beta\overline{\phi}|} 
\rho_{\phi}\left(\frac{\varepsilon}{\beta\overline{\phi}}\right)
~.  
\label{e2_3_2_2}
\end{equation}
If the coefficient $\alpha_{22}$ does not depend on $N$ then 
$\beta \propto N$ and the location of the main band 
$\langle g_0(\varepsilon)\rangle$ also scales with $N$. 
In the opposite case of $\beta$ being independent of $N$, 
the ensemble-averaged spectral density is $N$-independent. 
Fig.~\ref{f6} illustrates the high quality of the approximate expression 
for $\langle g_0(\varepsilon)\rangle \simeq \langle g(\varepsilon)\rangle$ 
in the band region  
(the solid line in Fig.~\ref{f6}) by comparison with the results 
obtained by direct diagonalization of the Hamiltonian matrix 
(the dashed line in Fig.~\ref{f6}).  

Similarly to the electronic problem, 
the contribution of $\langle \delta g(\varepsilon)\rangle$ to the spectral 
density is negligible ($\propto 1/N$) in the main band region. 
A specific feature of the vibrational problem is that one of the peak-shaped 
contributions from  $\langle \delta g(\varepsilon)\rangle$ is always 
of the $\delta$-functional form at exactly zero energy. 
For the multiplicative interactions, 
this is the only peak-shaped contribution, i.e.  
$\langle \delta g(\varepsilon)\rangle = N^{-1}\delta(\varepsilon)$. 
This is a consequence of the global translational invariance of the 
Hamiltonian and also follows from the solution 
of the equation $\widetilde{D}(\varepsilon_*)=0$, which reads 
\begin{equation}
 \widetilde{D}(\varepsilon_*) = 1+ 
\diagup\hskip-15pt\int\limits_{-\infty}^{\infty}
\frac{\beta \phi_i^2}{\varepsilon_*-\beta\overline{\phi}\phi_i}
\rho_{\phi}(\phi_i)\text{d}\phi_i =0 
~.  
\label{e2_3_2_3}
\end{equation}
Obviously, the value of $\varepsilon_*=0$ is the solution of 
Eq.~(\ref{e2_3_2_3}) and thus the peak is located at zero energy. 
This statement holds for an arbitrary realization of disorder, when 
the integrals in Eq.~(\ref{e2_3_2_3}) should be replaced by finite sums 
and thus the peak associated with $\varepsilon_*$ is 
of $\delta$-functional form.  

Therefore, the vibrational problem defined on the FCG with multiplicative 
interactions has a relatively simple ensemble-averaged spectral density. 
It contains the main band which is obtained by rescaling of 
the probability distribution for interaction parameters 
and a $\delta$-functional peak at zero energy. 

The results obtained for the vibrational problem can be applied to the  
investigation of the transport properties of communication networks 
characterized by symmetric transmission rates $W_{ij}=-V_{ij} \ge 0$ 
(i.e. $\beta < 0$) and  $W_{ij}=W_{ji}$. 
For example, the above analysis for separable node-node interactions 
is applicable for a network with communication channels charcterized 
by multiplicative functions \cite{Guimera_02}. 
In this case, the dynamical characteristics of a packet propagating 
through the network crucially depend on the functional form of 
the distribution, $\rho_\phi(\phi_i)$, of the interaction parameters $\phi_i$. 
One of these characteristics is the return probability, 
$\langle P_0(t)\rangle$, 
of the packet to the starting point of its diffusion through the 
network (see e.g. \cite{Friedrichs_88}), 
\begin{equation}
\langle P_0(t) \rangle = N^{-1}\text{Tr}\,\exp\{\hat{\bf H}t\}=
\int\limits_{-\infty}^\infty e^{\varepsilon t}
\langle g_0(\varepsilon) \rangle \text{d}\varepsilon + \frac{1}{N}
~, 
\label{e2_3_2_4}
\end{equation}
where $\langle g_0(\varepsilon) \rangle$ is given by Eq.~(\ref{e2_3_2_2}). 
The spectrum of the Hamiltonian~(\ref{e1_1}) is negative except 
for  one 
eigenvalue located exactly at zero which gives rise to 
the last term in Eq.~(\ref{e2_3_2_4}) describing the random return of the 
packet to the origin. 
The time dependence of the return probability is dictated by the 
functional form of $\rho_\phi(\phi)$ in Eq.~(\ref{e2_3_2_2}). 
If the distribution $\rho_\phi(\phi)$ has the form of a band, 
$\phi \in \left[\phi_{\text{min}},\phi_{\text{max}} \right]$, 
separated from zero by a gap, i.e. $\phi_{\text{min}} > 0$, then 
the long-time behaviour of the first term in Eq.~(\ref{e2_3_2_4})
for the  return probability is exponential, 
$\langle P_0(t) \rangle 
\propto \exp(-\beta\overline{\phi}\phi_{\text{min}} t)$. 
On the other hand, if the distribution of $\phi$ starts from zero with 
$\rho_{\phi}(\phi\to 0) \propto \phi^{\alpha}$ with $\alpha > -1$, 
the return probability decays with time according to the power law, 
$\langle P_0(t) \rangle 
\propto t^{-\alpha -1}$ for large times. 
This property can be quite important for monitoring the localization 
properties of the information packets by means of choosing the 
appropriate distribution for the interaction parameters $\phi_i$.

\subsection{\label{s2_3_3} 
Polynomial interactions of higher orders} 

To conclude this section about polynomial interactions defined on the FCG,  
we should emphasise that there are no principal difficulties 
in extending the above analysis to the cases of higher-order interactions 
with $n>2$. 
The features of the ensemble-averaged spectral density will be the same 
as for the multiplicative interactions. 
The shape of the main band can vary significantly depending on the 
order of interaction and the probability distributions of the bare energies 
and interaction parameters. 

\begin{figure}[ht] %
\vskip2truecm 
\centerline{\includegraphics[width=13cm]{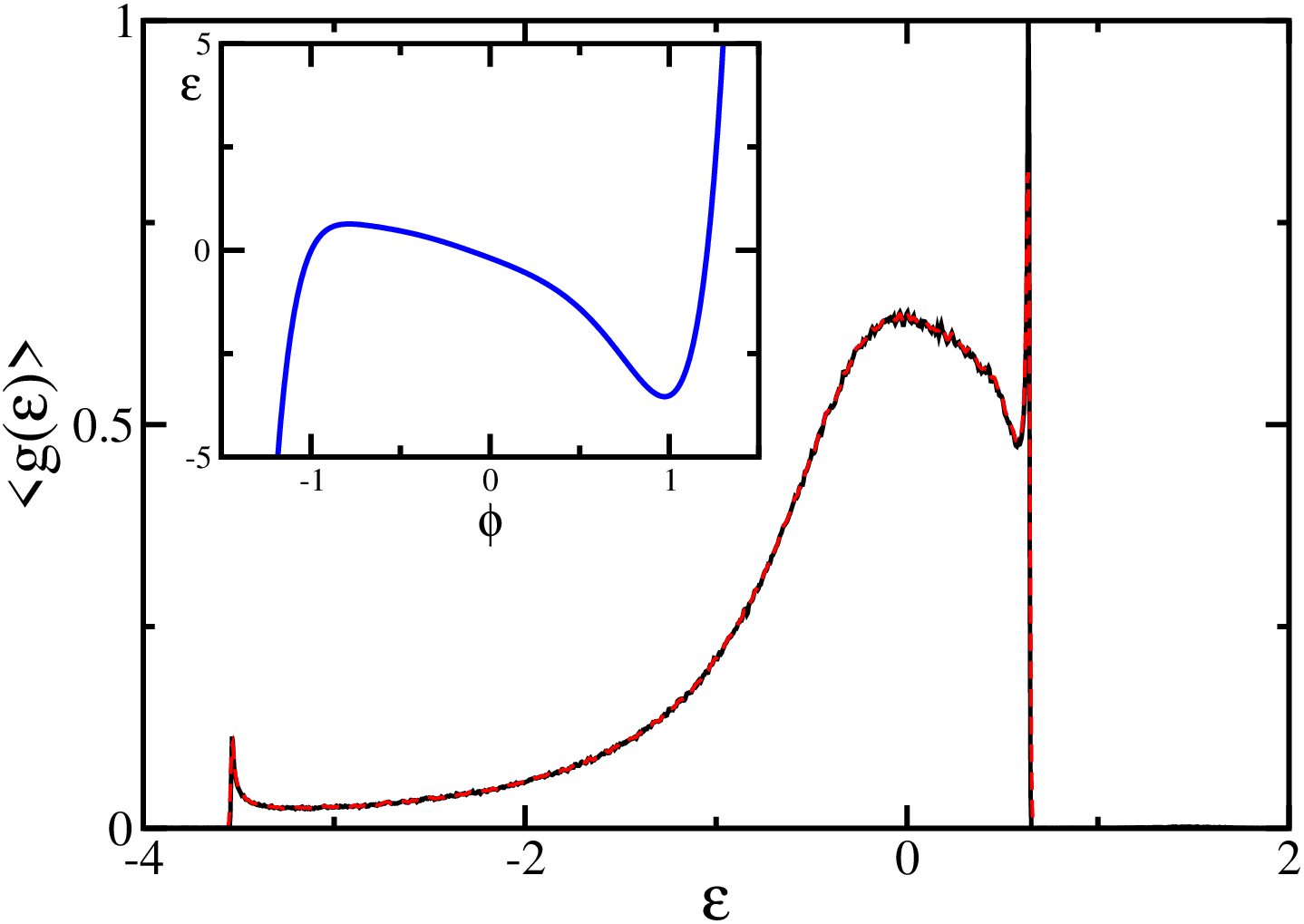}}  
\caption{(Color online) 
The exact (solid line) and approximate (the dashed line) 
ensemble-averaged DOS (over $10^3$ realizations) for the FCG ($N=3000$) 
with  node-node interactions defined 
by Eq.~(\ref{e1_2}) in which $n=6$. 
The elements of the $\bm{\alpha}$-matrix are random values 
withdrawn from the uniform distribution defined in the 
interval, $\alpha_{ij}\in [-1,1]$. 
The on-site energies and interaction parameters are  
distributed according to $\rho_\epsilon(\epsilon)=\delta(\epsilon)$ and 
$\rho_\phi(\phi)=N(0,0.16)$, respectively. 
The non-linear $\phi-\varepsilon$ map is shown in the inset. 
  }  
\label{f7} 
\end{figure}

A typical example is shown in Fig.~\ref{f7} for 
the electronic problem ($\gamma=0$) defined on the 
FCG with polynomial interactions 
of order $n=6$ characterized by a normal distribution of 
the interaction parameters and a $\delta$-functional distribution of the 
bare energies. 
In this case, the convolution of the probability distributions 
in Eq.~(\ref{e2_7}) is trivial and the band shape is dictated by 
the non-linear $\phi-\varepsilon$ map (see the inset in Fig.~\ref{f7})  
given by the polynomial of  order $2n-1$  with 
the singularities in the DOS being due to the extremal points 
in the $\phi-\varepsilon$ map. 
The ensemble-averaged DOS calculated according to 
Eq.~(\ref{e2_7}) (the dashed line) is practically identical to the 
exact one (the solid line) obtained by direct diagonalization 
of the Hamiltonian. 
The isolated levels, $\varepsilon_{i*}$, 
can be found if necessary by solving the equation for the 
real part of the spectral determinant, $\widetilde{D}(\varepsilon_{i*})=0$. 

\section{\label{s4} Limitations} 

In the previous section, we have demonstrated that the evaluation 
of the spectral density and its  configurational 
averaging for the Hamiltonian defined on the FCG 
 can be performed analytically in some special cases. 
Namely, this can be done  for the 
binary FCG exactly and approximately for the FCG with 
a particular polynomial type of the node-node interactions. 
One of the restrictions for the polynomial interactions is that 
the order of interactions must be much less than the number of 
nodes in the FCG, $n\ll N$. 
Generally speaking the solution given by Eq.~(\ref{e2_7}) is valid 
for arbitrary value of $n$ but it becomes ''useless'' for $n\agt N$ 
in the sense that the conclusions made above about the structure 
of $ \langle g(\varepsilon)\rangle $ do not necessarily hold for this case. 
For example, the maximum number of levels split from the main 
band due to polynomial interactions should be less or equal to the order of 
interactions, $n$, and therefore it can reach the value of $N$ if $n\agt N$. 
This means  
that the main band described by $ \langle g_0(\varepsilon)\rangle $ in 
Eq.~(\ref{e2_7}) can disappear completely and a new band or set of levels 
arises due to the contribution from $ \langle \delta g(\varepsilon)\rangle$ 
(see an example below). 
The positions of levels split from the main band should be found by 
solving the $n$-th order polynomial for the real part of the spectral 
determinant. 
The complexity of this problem is not less than that of 
the original eigenproblem 
and thus solution~(\ref{e2_7}) becomes not very informative. 

The condition $n\ll N$ can be broken for a very important type 
of the node-node interactions, 
$V_{ij}=V(|\phi_i-\phi_j|)$, depending on Euclidean distance, 
$|\phi_i-\phi_j|$, between nodes, where 
the $\phi_i$ play the role of the node coordinates which can be random 
values 
(for simplicity, we analyse one-dimensional space). 
For example, let us consider a set of nodes randomly 
displaced from the sites in the ideal linear chain, 
so that 
($E[\phi_j]= ja$; $a$ is the mean distance between 
nearest sites and $j$ is an integer). 
Assume also that the node-node interactions  decay exponentially
 with the distance, 
\begin{equation}
 V_{ij}=V_0 \exp\{-[\beta(\phi_i-\phi_j)]^2\} 
 ~,   
\label{e4_1}
\end{equation}
where $\beta$ is the inverse interaction length and $V_0$ is a constant.  
The interactions $V_{ij}$ given by Eq.~(\ref{e4_1}) can be expanded in a  
Taylor series generally containing an infinite number of terms  
and thus presented in the form of Eq.~(\ref{e1_2}) with $n \gg N$.  

However, in  the case of long-range interactions when 
the typical interaction length, $\beta$, is comparable to the system 
size, $L$, i.e. $\beta^{-1} \agt L=Na$, the Taylor series for $V_{ij}$ 
contains only a finite number of terms, $n\ll N$, and the 
interactions are of the polynomial type. 
Consequently, the ensemble-averaged spectral density should be 
well approximated in the main band region by the function 
$\langle g_0(\varepsilon)\rangle$ evaluated according to 
Eq.~(\ref{e2_7}) which gives for the electronic problem, 
\begin{equation} 
\langle g_0(\varepsilon)\rangle=\rho_\epsilon(\varepsilon -V_0)
 ~,   
\label{e4_2}
\end{equation}
because $\tilde{\epsilon}_i = \epsilon_i + V_{ii}=\epsilon_i +V_0$.  
It follows from Eq.~(\ref{e4_2}) that $\langle g_0(\varepsilon)\rangle$ 
does not depend on the distribution of $\phi_i$ at all. 
Indeed, we have found numerically  that in the case of long-range interactions 
 the ensemble-averaged spectral density follows the theoretical 
prediction~(\ref{e4_2}) 
(the black curve in Fig.~\ref{f7} obtained by direct diagonalization 
for $\beta N a =1$  
is indistinguishable from that obtained using Eq.~(\ref{e4_2})). 
In this regime,  all the 
nodes interact with each other at approximately the same strength, 
$V_{ij}\simeq V_0$,  and the system is equivalent to the FCG with 
on-site energy disorder only. 

\begin{figure}[ht] %
\vskip2truecm 
\centerline{\includegraphics[width=13cm]{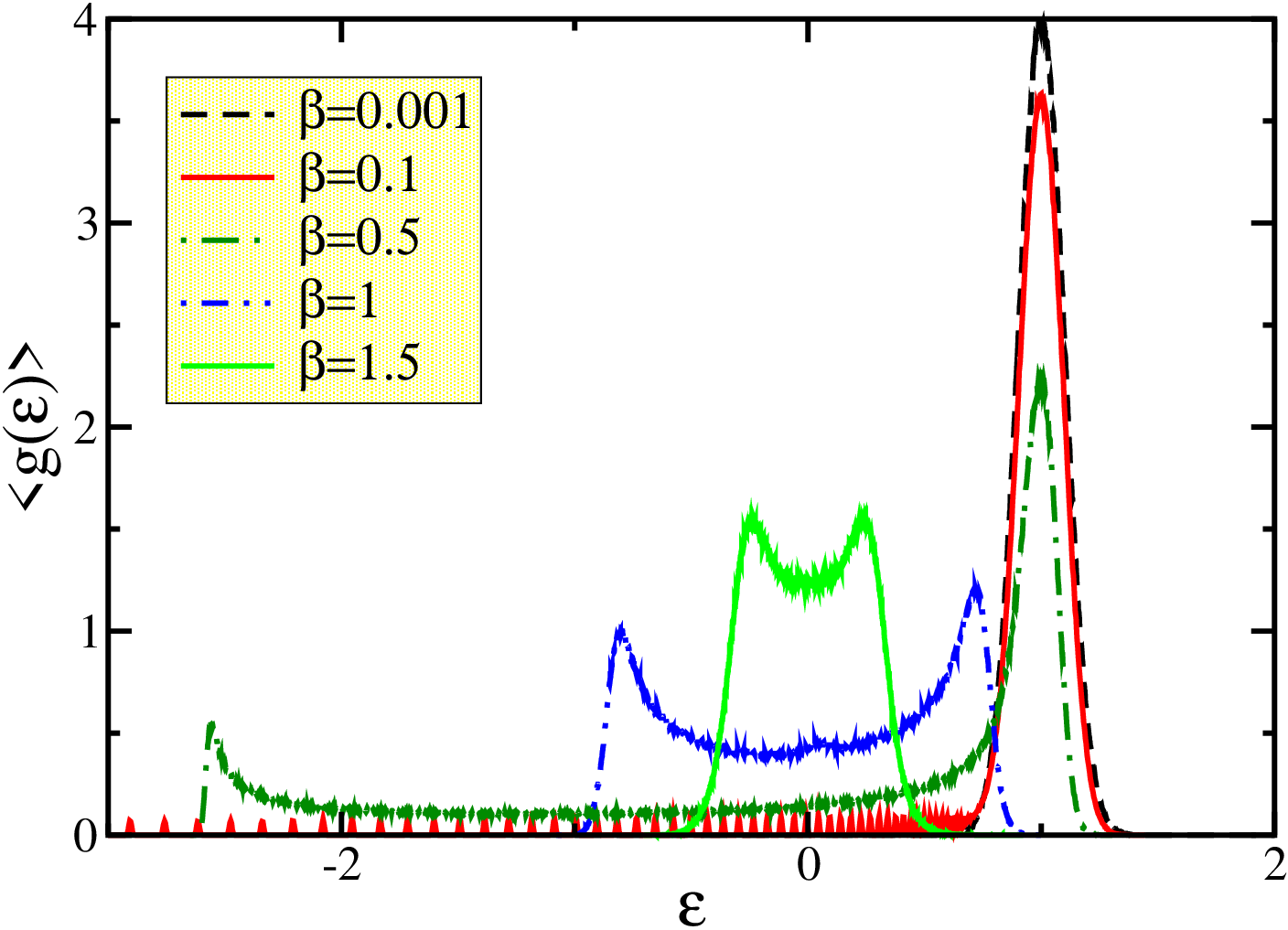}}  
\caption{(Color online) 
The exact ensemble-averaged DOS (over $3\times 10^3$ configurations) 
for a linear chain on $N=10^3$ nodes interacting with each other 
according to Eq.~(\ref{e4_1}) ($V_0=1$ and $a=1$) for different 
values of the inverse interaction length $\beta$ (as marked). 
The on-site energies and interaction parameters are normally 
distributed according to $\rho_\epsilon(\epsilon)=N(0,0.01)$ and 
$\rho_\phi(\phi)=N(1,0.01)$, respectively. 
  }  
\label{f8} 
\end{figure}

When the typical interaction length decreases, more terms in the 
Taylor series for $V_{ij}$ should be kept for the accurate representation 
of function~(\ref{e4_1}) leading to an increase in the order of the polynomial 
interactions. 
An increase in the value of $n$ gives rise to more and more levels split down 
off the main band (see the red solid curve in Fig.~\ref{f8} for $\beta  N a= 10^2$). 
These separate levels eventually form a broad band for medium-range 
interactions (see the dark green dashed curve in  Fig.~\ref{f8} 
for $\beta  N a= 5\times 10^2$) which 
transforms to a relatively narrow band  of width, 
$\Delta \simeq 4V_0 \exp(-\beta^2a^2)$, having  
a well-recognizable shape of a band for the spectrum of 
an ideal linear chain with nearest-neighbour interactions 
 broadened by diagonal and off-diagonal disorder 
(see the blue double-dot-dashed and light green solid curves  in Fig.~\ref{f8} 
for $\beta  N a=  10^3$ and 
$\beta  N a= 1.5\times 10^3$, respectively).  
The original main band centred around 
$\varepsilon \simeq \overline{\varepsilon}+ V_0$ for long-range interactions 
eventually disappears with increasing value of $\beta$.

Therefore, we can conclude that the theory for polynomial interactions 
presented in Sec.~\ref{s3} can be applied to systems with Euclidean 
long-range interactions but fails to describe the short-range interactions.  

The other limitation of the above analysis concerns its applicability 
to  the disordered complex networks of the FCG topology only. 
Of course, the topology of real complex networks is much more 
complicated (e.g. scale-free or small-world topologies 
\cite{Dorogovtsev_03:book}) and the FCG can be considered as a first 
approximation for the networks with high node-node connectivity.

\section{\label{s5} Conclusions}

To conclude, we have analysed the spectral properties of the 
Hamiltonian both for electronic and vibrational problems defined 
on the fully-connected graphs with a special type (polynomial) 
of interactions. 
 
Our main finding is the analytical formula (see Eq.~\ref{e2_7}) 
for the ensemble-averaged spectral density. 
The ensemble-averaged spectral density has two contributions 
with clear physical 
interpretations: (i) the first contribution describes the main spectral band 
and (ii) the second one is related to the set of discrete levels separated 
from the band. 
The main band originates from the bare spectral band shifted and deformed 
by means of a convolution of two probability distributions of bare energies 
and interaction parameters. 
The discrete levels are split from the bare spectral band due to interactions 
and the number of such levels depends on the order of the interactions. 

The approximate analytical configurational averaging is possible due to 
the availability of the exact analytical solution for the resolvent matrix 
elements for a particular realization of disorder (see Appendix~\ref{app_a}). 
Technically, configurational averaging is done with the use of the 
central limit theorem by replacing the function of fluctuating macroscopic 
sums by the function of the means of these sums (see Eq.~\ref{e2_4}). 
This step can be justified and illustrated by a more rigorous approach for a 
particular case (see Appendix~\ref{app_b}). 
All the final results are convincingly supported by numerics.

Both the electronic and vibrational problems discussed in the paper 
can be mapped to 
the contact process in the dilute regime and the stochastic diffusion problem, 
respectively. 
The results for the electronic case can thus be used for obtaining  
the lower bound estimate of the critical parameter for the contact 
process while the results for the vibrational case allow the return 
probability for an information packet to be evaluated for the communication 
networks.

\appendix
\section{Exact solution for the resolvent matrix elements}
\label{app_a}

The resolvent matrix elements for the Hamiltonian given by Eq.~(\ref{e1_1}) 
can be found exactly for the node-node interactions of 
polynomial type ~(\ref{e1_2}). 
In order to demonstrate this let us recast the equation for the resolvent, 
$(\varepsilon {\hat{\bf I}} - \hat{\bf H})\hat{\bf G} = {\hat{\bf I}}$, 
as 
\begin{equation}
G_{ij}=\frac{\delta_{ij}}{\varepsilon -\tilde\epsilon_i}- 
\frac{1}{\varepsilon -\tilde\epsilon_i}
\sum_k  \bm{\psi}_i^T \bm{\varphi}_k G_{kj} 
~, 
\label{app_a_1}
\end{equation}
where the renormalized bare energy, $\tilde\epsilon_i$, is 
\begin{equation}
\tilde\epsilon_i= \epsilon_i + \gamma 
 \bm{\psi}_i^T\sum_j^N \bm{\varphi}_j 
+(1-\gamma)\bm{\psi}_i^T\bm{\varphi}_i 
= \epsilon_i + \widetilde{V}_{ii}
~,  
\label{app_a_2}
\end{equation}
with $\bm{\psi}_i^T=\bm{\varphi}_i^T \bm{\alpha}$. 
The value of $\widetilde{V}_{ii}=\gamma 
 \bm{\psi}_i^T\sum_j^N \bm{\varphi}_j 
+(1-\gamma)\bm{\psi}_i^T\bm{\varphi}_i 
= \epsilon_i + \widetilde{V}_{ii}$ 
generally depends on all $\phi_j$ but 
$\widetilde{V}_{ii}=V_{ii}(\phi_i)$ for the electronic problem ($\gamma=0$).

The node-separable type of polynomial interactions in 
Eq.~(\ref{e1_2}) ($i-j$ interaction is proportional to the matrix product 
of separate node characteristics)  allows 
factorization to be performed there by introducing 
$\bf{x}_j =\sum_k \bm{\varphi}_k G_{kj}$, so that 
\begin{equation}
G_{ij}=\frac{\delta_{ij}}{\varepsilon -\tilde\epsilon_i}- 
\frac{1}{\varepsilon -\tilde\epsilon_i}
\bm{\psi}_i^T\bf{x}_j  
~.  
\label{app_a_3}
\end{equation}
Eq.~(\ref{app_a_1}) can be multiplied by $\bm{\varphi}_i$ and summed over 
$i$ and thus transformed to equation
%
%
%
%
which can be solved for $\bf{x}_j$, 
\begin{equation}
\bf{x}_j=\left(\bf{I}+  
\sum_i^N \frac{\bm{\varphi}_i\bm{\psi}_i^T}{\varepsilon -\tilde\epsilon_i}
\right)^{-1}  
\frac{\bm{\varphi}_j}{\varepsilon -\tilde\epsilon_j} 
~,  
\label{app_a_4}
\end{equation}
Substitution of $\bf{x}_j$ into Eq.~(\ref{app_a_3}) results in the 
required result for the resolvent matrix elements, 
\begin{equation}
G_{ij}=\frac{\delta_{ij}}{\varepsilon -\tilde\epsilon_i}- 
\frac{\bm{\psi}_i^T}{\varepsilon -\tilde\epsilon_i}
\left(\bf{I}+  
\sum_k^N \frac{\bm{\varphi}_k \bm{\psi}_k^T}{\varepsilon -\tilde\epsilon_k}
\right)^{-1}
\frac{\bm{\varphi}_j}{\varepsilon -\tilde\epsilon_j}
~.  
\label{app_a_5}
\end{equation}

The DOS is expressed via $\text{Tr}\hat{\bf G}$, which is given by the 
following expression, 
\begin{equation}
\text{Tr}\hat{\bf G}=
\sum_i^N \frac{1}{\varepsilon -\tilde\epsilon_i}- 
\sum_i^N 
\frac{\bm{\psi}_i^T}{\varepsilon -\tilde\epsilon_i}
{\bf D}^{-1}
\frac{\bm{\varphi}_i}{\varepsilon -\tilde\epsilon_i}
~,  
\label{app_a_6}
\end{equation}
and 
\begin{equation}
{\bf D}= \left(\bf{I}+  
\sum_k^N \frac{{\bm{\varphi}}_k{\bm{\psi}}_k^{T}}
{\varepsilon -\tilde\epsilon_k}\right)
~,  
\label{app_a_7}
\end{equation}
The eigenvalues, $\varepsilon_i$, 
of the Hamiltonian coincide with the poles of the 
resolvent which are the roots of the spectral determinant, 
 $D(\varepsilon)=\text{det}\,{\bf D}(\varepsilon)|$, i.e. 
$D(\varepsilon_i)=0$. 
This follows from  the form of Eq.~(\ref{app_a_6}) in which the contributions 
to the denominator from the first sum,  
$\propto (\varepsilon -\tilde\epsilon_i)$, are cancelled by the similar 
terms in Eq.~(\ref{app_a_6}). 

It is easy to show that Eq.~(\ref{app_a_6}) can be recast in an elegant 
form via the derivative of the spectral determinant, 
\begin{equation}
\text{Tr}\hat{\bf G}=
\sum_i^N \frac{1}{\varepsilon -\tilde\epsilon_i}+
\frac{\text{d}\ln[\text{det}({\bf D})]}{\text{d}\varepsilon}
~.  
\label{app_a_8}
\end{equation}
Indeed, rewriting the second term from Eq.~(\ref{app_a_6}) 
in the following form, 
\begin{equation}
\sum_i^N 
\frac{\bm{\psi}_i^T}{\varepsilon -\tilde\epsilon_i}
{\hat{\bf D}}^{-1}
\frac{\bm{\varphi}_i}{\varepsilon -\tilde\epsilon_i}=  
[\text{det}({\bf D})]^{-1}\sum_{i}^N\sum_{k,m}^n 
\frac{(\bm{\psi}_i^T)_k}{\varepsilon -\tilde\epsilon_i} 
C_{mk} 
\frac{(\bm{\varphi}_i)_m}{\varepsilon -\tilde\epsilon_i}
~,  
\label{app_a_9}
\end{equation}
where ${\bf C}$ stands for the matrix of cofactors for matrix ${\bf D}$, 
and comparing it with the derivative of $\text{det}({\bf D})$, 
\begin{equation}
\frac{\text{d}\,\text{det}({\bf D})}{\text{d}\varepsilon} = 
\sum_{m,k} \frac{\text{d} D_{mk}}{\text{d}\varepsilon} C_{mk} = 
-\sum_{m,k} \frac{(\bm{\varphi}_i)_m(\bm{\psi}_i^T)_k}{(\varepsilon -\tilde\epsilon_i)^2} C_{mk}
~,  
\label{app_a_10}
\end{equation}
we arrive at Eq.~(\ref{app_a_8}). 
Finally, taking the imaginary part of Eq.~(\ref{app_a_8}) and using then 
Eq.~(\ref{e1_4}) leads to Eq.~(\ref{e2_2}).

\section{\label{app_b} 
Configurational averaging by direct integration}
In this Appendix, we give an alternative derivation of the expression 
for the ensemble-averaged spectral density in the case of 
the multiplicative node-node interaction, $V_{ij}=\alpha_{22}\phi_i \phi_j$, 
for the random Hamiltonian defined on the FCG. 
The derivation is similar in some aspects to that given in 
Ref.~\cite{Bogomolny_01} for the mean density of eigenvalues of a 
2D integrable billiard. 

The starting point for the derivation is Eq.~(\ref{app_a_6}) 
recasted for the multiplicative interaction~(\ref{e2_3_1_1}) in the 
following form:
\begin{eqnarray}
\text{Tr}\hat{\bf G}&=&
\sum_i^N \frac{1}{\varepsilon -\tilde\epsilon_i}
\,-\,  
\sum_i^N \frac{\alpha_{22}\phi_i^2}{(\varepsilon - \tilde{\epsilon}_i)^2}
\left(1+ 
\sum_i^N \frac{\alpha_{22}\phi_i^2}{\varepsilon - \tilde{\epsilon}_i}
 \right)^{-1} 
\nonumber 
\\
&=& \sum_i^N \frac{1}{\varepsilon -\tilde\epsilon_i}
\,-\, \text{i}\sum_i^N \frac{\alpha_{22}\phi_i^2}{(\varepsilon - \tilde{\epsilon}_i)^2} 
\int\limits_{-\infty}^{0-}e^{\text{i}k
\left(1+ 
\sum_i^N \frac{\alpha_{22}\phi_i^2}{\varepsilon - \tilde{\epsilon}_i} \right)} 
\text{d}k
~,  
\label{app_b_1}
\end{eqnarray}
where 
\begin{equation}
\tilde{\epsilon_i}=\epsilon_i+\alpha_{22}\phi_i^2
~, 
\label{app_b_2}
\end{equation}
for the electronic 
problem ($\gamma = 0$) analysed below for concreteness (the analysis 
can be easily extended to the vibrational problem ($\gamma=1$)).    
Using definition~(\ref{e1_4}) and Eq.~(\ref{e1_3}) we obtain 
the expression for $\langle g(\varepsilon)\rangle =
 \langle g_0(\varepsilon)\rangle + \langle \delta g(\varepsilon)\rangle $, 
where $\langle g_0(\varepsilon)\rangle$ coincides with the first integral  
term in Eq.~(\ref{e2_3}) with $V_{ii}=\alpha_{22}\phi_i^2$ 
and $\langle \delta g(\varepsilon)\rangle $ is 
given by
\begin{equation}
\langle  \delta g(\varepsilon)\rangle = \frac{1}{\pi N}\, \text{Re}\, 
\int\limits_{-\infty}^{0-}e^{\text{i}k}\text{d}k 
\idotsint 
\sum_j^N \frac{\alpha_{22}\phi_j^2}{(\varepsilon - \tilde{\epsilon}_j)^2} 
\prod_i 
e^{\text{i}k\frac{\alpha_{22}\phi_i^2}{\varepsilon - \tilde{\epsilon}_i}}
\rho_{\phi}(\phi_i) \rho_{\epsilon}(\epsilon_i) 
\text{d}\phi_i\text{d}\epsilon_i
~,
\label{app_b_3}
\end{equation}
or by the equivalent expression 
(assuming for definiteness that $\alpha_{22}>0$), 
\begin{equation}
\langle  \delta g(\varepsilon)\rangle = \frac{1}{\pi}
 \text{Re}\, 
\int\limits_{-\infty}^{0-}e^{\text{i}k}\left[F(k,\varepsilon)\right]^{N-1} 
Q(k,\varepsilon)
\text{d}k 
~,
\label{app_b_4}
\end{equation}
with 
\begin{equation}
F(k,\varepsilon) = \int\limits_{-\infty}^{\infty} \rho_{\phi}(\phi_i)
f(\phi_i,k,\varepsilon)\text{d}\phi_i = 
\int\limits_{-\infty}^{\infty} \rho_{\phi}(\phi_i)
\int\limits_{-\infty}^{\infty} 
e^{\text{i}k\frac{\alpha_{22}\phi_i^2}{\varepsilon - \tilde{\epsilon}_i}}
\rho_{\epsilon}(\epsilon_i) \text{d}\epsilon_i
\text{d}\phi_i
~,
\label{app_b_5}
\end{equation}
and 
\begin{equation}
Q(k,\varepsilon) = \int\limits_{-\infty}^{\infty} \rho_{\phi}(\phi_i)
q(\phi_i,k,\varepsilon)\text{d}\phi_i = 
\int\limits_{-\infty}^{\infty} \rho_{\phi}(\phi_i)
\int\limits_{-\infty}^{\infty} 
\frac{\alpha_{22}\phi_i^2}{(\varepsilon - \tilde{\epsilon}_i)^2}
e^{\text{i}k\frac{\alpha_{22}\phi_i^2}{\varepsilon - \tilde{\epsilon}_i}}
\rho_{\epsilon}(\epsilon_i) \text{d}\epsilon_i
\text{d}\phi_i
~,
\label{app_b_6}
\end{equation}
where the function $q(\phi_i,k,\varepsilon)$ is related to 
$f(\phi_i,k,\varepsilon)$ by the following equation, 
\begin{equation}
q(\phi_i,k,\varepsilon) = -\frac{1}{\alpha_{22}\phi_i^2}\, 
\frac{\partial^2 f(\phi_i,k,\varepsilon)}{\partial k^2}
~. 
\label{app_b_7}
\end{equation}

The next step is in the evaluation of the integral, 
\begin{equation}
f(\phi_i,k,\varepsilon) = 
\int\limits_{-\infty}^{\infty} 
e^{-\text{i}\frac{\chi}
{z -\epsilon_i}}
\rho_{\epsilon}(\epsilon_i) \text{d}\epsilon_i
~, 
\label{app_b_8}
\end{equation}
($\chi \equiv -k\alpha_{22}\phi_i^2 > 0$) 
with  the integrand exhibiting  an essential singular point at 
$\epsilon_i=\varepsilon +\text{i}0 - \alpha_{22}\phi_i^2 \equiv z$. 
This can be done by expanding the exponential function in a Taylor 
series and integrating each term, 
\begin{eqnarray}
f &=& 
\int\limits_{-\infty}^{\infty} 
\rho_{\epsilon}(\epsilon_i) 
\left[
1-\text{i}\chi\,\frac{1}{z-\epsilon_i} + 
\frac{(\text{i}\chi)^2}{2}\,\frac{1}{(z-\epsilon_i)^2}+\cdots 
\right]
\text{d}\epsilon_i 
\nonumber 
\\
&=& 1-\text{i}\chi\left(R(z)+\text{i}I(z)\right)+ 
\frac{\chi^2}{2}\left(R'(z)+\text{i}I'(z)\right)+O(\chi^3/\Delta^3)
~, 
\label{app_b_9}
\end{eqnarray}
where the functions $R(z)$ and $I(z)$ are defined by Eq.~(\ref{e2_6}). 
In order to evaluate the third term in Eq.~(\ref{app_b_9}) we integrated 
once by parts and used the following identity, 
\begin{equation}
R'(z)=  
\diagup\hskip-15pt\int\limits_{-\infty}^{\infty}
\frac{\rho_{\epsilon}(z) - \rho_{\epsilon}(\epsilon_i)}
{(z-\epsilon_i)^2}
 \text{d}\epsilon_i = 
\diagup\hskip-15pt\int\limits_{-\infty}^{\infty}
\frac{ \rho_{\epsilon}'(\epsilon_i)}
{z-\epsilon_i}
 \text{d}\epsilon_i 
~.  
\label{app_b_10}
\end{equation}
In expansion~(\ref{app_b_9}), we keep only the terms  up
 to the second order in $\beta/\Delta$ including because this is 
enough for obtaining the leading term for $q(\phi_i,k,\varepsilon)$ 
according to Eq.~(\ref{app_b_7})
\begin{equation}
q(\phi_i,k,\varepsilon) = -\alpha_{22}\phi_i^2 
\left[
R'(\varepsilon-\alpha_{22}\phi_i^2) + 
\text{i} I'(\varepsilon-\alpha_{22}\phi_i^2)
\right]
~.  
\label{app_b_11}
\end{equation}
The higher order terms in $\beta/\Delta$ are small both for
$f(\phi_i,k,\varepsilon)$ and $q(\phi_i,k,\varepsilon)$
because typical values 
of $|k|$ significantly contributing into integral~(\ref{app_b_4}) 
in the band region are 
$|k|\sim \Delta/(\alpha_{22}\phi_i^2 N)$ (see below) and thus 
 $\beta/\Delta \sim 1/N \ll 1$. 
Note, that the same result for $f(\phi_i,k,\varepsilon)$ can be obtained 
by a different method based on the shift of the essential singularity to 
infinity as was suggested in Ref.~\cite{Bogomolny_01}. 

Substitution of Eqs.~(\ref{app_b_10})-(\ref{app_b_11}) 
into Eqs.~(\ref{app_b_5})-(\ref{app_b_6})
gives, 
\begin{eqnarray}
F(k,\varepsilon) &=& 
1+\text{i}k \left[R_1(\varepsilon) + \text{i} I_1(\varepsilon)\right] 
+ \frac{k^2}{2}
\left[R_2'(\varepsilon) + \text{i} I_2'(\varepsilon)\right]
\label{app_b_12}
\\
Q(k,\varepsilon) &=& 
-R_1'(\varepsilon) - \text{i} I_1'(\varepsilon)
~,
\label{app_b_13}
\end{eqnarray}
where
\begin{eqnarray}
R_m(k,\varepsilon) &=& \int\limits_{-\infty}^{\infty} \rho_{\phi}(\phi_i)
\left[\alpha_{22}\phi_i^2 \right]^m 
R(\varepsilon-\alpha_{22}\phi_i^2) \text{d}\phi_i
\label{app_b_14}
\\
I_m(k,\varepsilon) &=& \int\limits_{-\infty}^{\infty} \rho_{\phi}(\phi_i)
\left[\alpha_{22}\phi_i^2 \right]^m 
I(\varepsilon-\alpha_{22}\phi_i^2) \text{d}\phi_i
~. 
\label{app_b_15}
\end{eqnarray}
Using the above expressions for $F(k,\varepsilon)$ and $Q(k,\varepsilon)$ 
we can rewrite Eq.~(\ref{app_b_4}) as 
\begin{equation}
\langle g(\varepsilon)\rangle = -\frac{1}{\pi}
 \text{Re}\left\{\left[R_1'(\varepsilon)+\text{i} I_1'(\varepsilon)\right] 
\int\limits_{-\infty}^{0-}e^{\psi(k)}
\text{d}k \right\} 
~,
\label{app_b_16}
\end{equation}
with
\begin{eqnarray}
&~&\psi(k) = \text{i}k+(N-1)\ln
\left[1+k(\text{i}R_1-I_1)+\frac{k^2}{2}(R_2'+\text{i}I_2' \right]
\nonumber
\\
&= & \text{i}k(1+NR_1) - k N I_1 +
\frac{k^2 N}{2} \left(R_1^2+R_2'-I_1^2\right) + 
\frac{\text{i}k^2 N}{2} \left(I_2'+2R_1I_1\right) + O(k^3) 
~.
\label{app_b_17}
\end{eqnarray}

There are two energy regions: (i) inside the band where  
$R_m(\varepsilon) \sim I_m(\varepsilon) 
\sim \alpha_{22}^m\overline{\phi^m}/\Delta$ and 
(ii) outside the band where $R_m(\varepsilon)
\sim \alpha_{22}^m\overline{\phi^m}/
|\varepsilon_* - \overline{\epsilon}|$ 
with $|\varepsilon_* - \overline{\epsilon}|\gg \Delta$ 
and $ I_m(\varepsilon)$ either approaches zero (e.g. exponentially for 
the normal distribution $\rho_{\epsilon}$) or identically equals zero for 
the box distribution. 
In these regions, integral~(\ref{app_b_16}) has different contributions 
to the total ensemble-averaged spectral density. 
Inside the band, we can ignore the terms $\propto k^2 $ 
in expression~(\ref{app_b_17}) for $\psi(k)$, so that $e^{\psi(k)}$ 
exponentially decays for $|k|\to \infty $ on the typical scale 
$k \sim 1/(NI_1)\sim \Delta/(\alpha_{22}\overline{\phi^2}N)$, and 
\begin{equation}
\langle \delta g(\varepsilon)\rangle = -\frac{1}{\pi N}
\frac{(1+NR_1)NI_1'- NR_1' NI_1}{(1+NR_1)^2 + (NI_1)^2}
~,
\label{app_b_18}
\end{equation}
which exactly coincides with the second term in Eq.~(\ref{e2_7}) 
bearing in mind that $\widetilde{D}=1+NR_1$ and 
$\widetilde{\widetilde{D}}=NI_1$. 

Outside the band, $\rho_\epsilon \to 0$ and thus $I_m \to 0$, so that 
the real linear term in $k$ in Eq.~(\ref{app_b_17}) becomes negligible 
and the next terms in the expansion must be kept, 
\begin{equation}
\psi(k) 
\simeq  \text{i}k(1+NR_1) +
\frac{k^2 N}{2} \left(R_1^2+R_2'\right) 
~,
\label{app_b_19}
\end{equation}
where 
\begin{equation}
R_2' \simeq - \iint \left[
\frac{\alpha_{22}\phi_i^2}{\varepsilon-\alpha_{22}\phi_i^2-\epsilon_i}
\right]^2 \rho_\phi(\phi_i) \rho_{\epsilon}(\epsilon_i) 
\text{d}\phi_i\, \text{d}\epsilon_i 
\equiv - \overline{x^2}
~,
\label{app_b_20}
\end{equation}
and $R_1=\overline{x}$, $x_i=\alpha_{22}\phi_i^2/
(\varepsilon-\alpha_{22}\phi_i^2-\epsilon_i)$. 
Note that Eq.~(\ref{app_b_19}) contains an extra term, $\propto R_1^2$, in 
comparison with a similar expression given in Ref.~\cite{Bogomolny_01} which 
is due to a more accurate expansion of $\psi$ in $k$. 
Straightforward evaluation of the Gaussian integral in Eq.~(\ref{app_b_18}) 
leads to 
\begin{equation}
\langle \delta g(\varepsilon)\rangle = 
-\frac{R_1'}{\sqrt{2\pi N \text{Var}[x_i]}}
e^{-\frac{(1+NR_1)^2}{2N\text{Var}[x_i]}}
=\frac{1}{N}
\frac{1}{\sqrt{2\pi\sigma^2}}
e^{-\frac{(\varepsilon-\varepsilon_*)^2}{2\sigma^2}}
~,
\label{app_b_21}
\end{equation}
with $\text{Var}[x_i]=-R_1^2-R_2'$ and 
\begin{equation}
\sigma^2 = 
- \frac{1}{N}\, \frac{R_1^2+R_2'}{(R_1')^2}
\simeq 
\frac{1}{N}\left( 
\text{Var}[\epsilon_i]+ \text{Var}[N\alpha_{22}\phi_i^2]
\right)
~. 
\label{app_b_22}
\end{equation}
The spectral density given by Eq.~(\ref{app_b_21}) represents a Gaussian peak 
of width $\sigma$ centred at $\varepsilon=\varepsilon_*$. 
The location of the peak 
$\varepsilon_* \sim \overline{\epsilon}-\alpha_{22}\overline{\phi^2}$ 
is the solution of the 
equation, $1+NR_1(\varepsilon_*)=0$ identical to Eq.~(\ref{e2_3_1_6}) and 
thus the expression for $\varepsilon_*$ coincides with that
given by Eq.~(\ref{e2_3_1_7}). 

All the derivations presented in this Appendix were undertaken under 
the assumption that the coefficient $\alpha_{22}$ does not depend on $N$. 
However, in the case when $\alpha_{22} \propto 1/N$ all the results 
obtained for the main band still hold but Eq.~(\ref{e2_3_1_7}) for 
the position of the separate levels is no longer correct in general and 
the equation, $1+NR_1(\varepsilon_*)=0$, should be solved without using 
the assumption that the level is well separated from the main band. 


\end{document}